\documentclass[10pt,aps,prc,twocolumn,superscriptaddress,floatfix,nofootinbib]{revtex4-1}
\usepackage{times}
\usepackage[T1]{fontenc}

\usepackage{amsmath,amssymb,mathrsfs}
\usepackage{bm}
\usepackage{ascmac}
\usepackage{geometry}
\usepackage{graphicx}

\def\0\\{\nonumber\\}
\def\bs#1{\boldsymbol{#1}}

\newcommand{\Ecm}[1]{$E_\mathrm{c.m.}$\,$=$\,{#1}\,MeV}

\newcommand{\beq}{\begin{equation}}
\newcommand{\eeq}{\end{equation}}
\newcommand{\beqn}{\begin{eqnarray}}
\newcommand{\eeqn}{\end{eqnarray}}

\makeatletter
\newcommand\footnoteref[1]{\protected@xdef\@thefnmark{\ref{#1}}\@footnotemark}
\makeatother

\begin{document}

\title{
Kinetic energy dissipation and fluctuations in strongly-damped heavy-ion collisions\\ within the stochastic mean-field approach
}

\author{Sakir Ayik}
\email[]{ayik@tntech.edu}
\affiliation{Physics Department, Tennessee Technological University, Cookeville, Tennessee 38505, USA}

\author{Kazuyuki Sekizawa}
\email[]{sekizawa@phys.sc.niigata-u.ac.jp}
\affiliation{Center for Transdisciplinary Research, Institute for Research Promotion, Niigata University, Niigata 950-2181, Japan}
\affiliation{Division of Nuclear Physics, Center for Computational Sciences, University of Tsukuba, Ibaraki 305-8577, Japan}

\date{December 23, 2020}

\begin{abstract}
\begin{description}
\item[Background]
Microscopic mean-field approaches have been successful in describing the most probable
reaction outcomes in low-energy heavy-ion reactions. However, those approaches are
known to severely underestimate dispersions of observables around the average values
that has limited their applicability. Recently it has been shown that a quantal transport
approach based on the stochastic mean-field (SMF) theory significantly improves the
description, while its application has been limited so far to fragment mass and charge dispersions.

\item[Purpose]
In this work, we extend the quantal transport approach based on the SMF theory
for relative kinetic energy dissipation and angular momentum transfer in
low-energy heavy-ion reactions.

\item[Methods]
Based on the SMF concept, analytical expressions are derived for the radial and
tangential friction and associated diffusion coefficients. Those quantal transport
coefficients are calculated microscopically in terms of single-particle orbitals within
the time-dependent Hartree-Fock (TDHF) approach.

\item[Results]
As the first application of the proposed formalism, we consider the radial linear
momentum dispersion, neglecting the coupling between radial and angular momenta.
We analyze the total kinetic energy (TKE) distribution of binary reaction products in
the $^{136}$Xe+$^{208}$Pb reaction at \Ecm{526} and compare with experimental
data. From time evolution of single-particle orbitals in TDHF, the radial diffusion coefficient
is computed on a microscopic basis, while a phenomenological treatment is introduced
for the radial friction coefficient. By solving the quantal diffusion equation for the radial
linear momentum, the dispersion of the radial linear momentum is obtained, from which
one can construct the TKE distribution. We find that the calculations provide a good
description of the TKE distribution for strongly-damped events with large energy losses,
$\mbox{TKEL}$\,$\gtrsim$\,150\,MeV. However, the calculations underestimate
the TKE distribution for smaller energy losses. Further studies are needed to improve
the technical details of calculations.

\item[Conclusions]
It has been shown that the quantal transport approach based on the SMF theory
provides a promising basis for the microscopic description of the TKE distribution
as well as the isotopic distributions in damped collisions of heavy ions at around
the Coulomb barrier.

\end{description}
\end{abstract}


\maketitle

\section{INTRODUCTION}

The nuclear dissipation plays a major role in nuclear dynamics such as heavy-ion
collisions as well as nuclear fission. In order to understand the nuclear dissipation
mechanism, a large amount of investigations have been carried out both experimentally
and theoretically over many years \cite{Ayik(1976),Agassi(1977),Randrup(1978),Randrup(1979)}.
In low-energy heavy-ion collisions at around the Coulomb barrier, the one-body
dissipation-fluctuation mechanism originating from nucleon exchange is essential.
The time-dependent Hartree-Fock (TDHF) approach provides a microscopic basis
for describing dissipative collisions at low energies. It incorporates with the one-body
dissipation mechanism and successfully describes the most probable dynamical path
of reaction dynamics \cite{Negele(review),Simenel(review:2012),Nakatsukasa(review),
Simenel(review:2018),Stevenson(2019),Sekizawa(2019),Simenel(2020)}. However, it is well known
that the mean-field treatment of the TDHF approach severely underestimates dynamical
fluctuations around the most probable path. Recent applications of the so-called
time-dependent random phase approximation (TDRPA), which is based on the
generalized variational principle of Balian and V\'en\'eroni \cite{BV(1981),Balian(1984),Balian(1985)},
provides a possible prescription for calculating dispersions of one-body observables in
low-energy heavy-ion reactions. The latter approach has been applied to calculate
mass and charge dispersions in heavy-ion collisions \cite{Broomfield(2009),Simenel(2011),
Williams(2018),Godbey(2019)}. Although there was an attempt to quantify kinetic
energy fluctuations in dissipative collisions in the past \cite{Marston(1985)}, its practical
applications are still scarce. This work is the first step toward the fully microscopic
description of dissipation and fluctuations of the relative motion of colliding nuclei
based on an alternative approach, the stochastic mean-field (SMF) theory
\cite{Ayik(2008)1,Lacroix(2014)1}.

It is crucially important to develop a microscopic basis for describing fluctuations
in the kinetic energy dissipation for providing a reliable prediction for producing
unknown unstable nuclei. In recent years, deep-inelastic collisions such as multinucleon
transfer and quasifission processes have engaged substantial interests, regarding
the possibility of efficient production of unknown neutron-rich heavy nuclei.
Production of transactinide nuclei in the superheavy region in deep-inelastic or
quasifission type processes in damped collisions of two heavy nuclei has been
explored. Besides, multinucleon transfer reactions at energies around the
Coulomb barrier are expected to be useful to produce neutron-rich heavy nuclei
along the neutron magic number $N$\,$=$\,126. (See, e.g., Refs.~\cite{Zhang(2018),
Adamian(2020)}, for recent reviews.) To provide a reliable prediction for production
of yet-unknown unstable nuclei, it is of paramount importance to properly describe
not only dispersions of mass and charge of reaction products, as was greatly improved
by the recent developments of the SMF approach \cite{Ayik(2016),Ayik(2017),
Ayik(2018)2,Yilmaz(2018),Ayik(2019)1,Ayik(2019)2,Yilmaz(2020),Sekizawa(2020)},
but also the distribution of dissipated relative kinetic energy during the collision.
The latter is directly connected with excitation energies of reaction products,
which should not be too large to maximize the production yield. Regarding the
ongoing worldwide experimental effort aiming at producing unknown neutron-rich
heavy nuclei \cite{Adamian(2020),Kozulin(2012),Barrett(2015),Vogt(2015),
Watanabe(2015),Welsh(2017),Desai(2019),Desai(2020),Desai(2020)2}, it is
an imperative task to develop a fully microscopic framework for dissipation and
fluctuations of the relative motion of colliding nuclei associated with nucleon exchange.

In this work, we develop a quantal transport formalism for dissipation and
fluctuations of the relative kinetic energy and the relative angular momentum
transfer based on the SMF approach. Analytical expressions for the radial and
tangential friction and associated diffusion coefficients are derived on the microscopic
basis. As a first step toward the fully microscopic description of energy and angular
momentum dissipation in low-energy heavy-ion reactions, in the present work,
we consider dissipation of the relative radial linear momentum, neglecting its
coupling with the angular momentum transfer. The kinetic energy dissipation
in the collision of $^{136}$Xe+$^{208}$Pb at \Ecm{526} is analyzed with
the newly developed approach and the total kinetic energy (TKE) distribution
is compared with the available experimental data \cite{Kozulin(2012)}.

The article is organized as follows. In Sec.~\ref{Sec:Formulation}, we present
derivation of the Langevin equations for the relative radial momentum and
the orbital angular momentum. In Sec.~\ref{Sec:TKE}, quantal expressions of
diffusion coefficients for the radial and angular momenta, and the joint probability
distribution function for these quantities are given. In Sec.~\ref{Sec:Results},
the numerical results of the TKE distribution for the $^{136}$Xe+$^{208}$Pb
reaction at \Ecm{526} are presented and compared with the experimental data.
A summary and conclusions are given in Sec.~\ref{Sec:Conclusions}.

\section{Fluctuation of the relative momenta within the SMF approach}\label{Sec:Formulation}

\subsection{Remarks on the SMF approach}

The SMF approach goes beyond the standard TDHF description and provides
a microscopic basis for describing the fluctuations around the most probable path
\cite{Ayik(2008)1,Lacroix(2014)1}. In the SMF approach, instead of a single
deterministic event in TDHF, an ensemble of mean-field events is considered,
which is associated with a distribution law. The single-particle density matrix
of an event $\lambda$ is given by
\beq
\rho^\lambda(\bs{r},\bs{r}',t) = \sum_{ij}\phi_j^*(\bs{r},t;\lambda)\rho_{ji}^\lambda\phi_i(\bs{r}',t;\lambda),
\eeq
where the wave functions in each event $\lambda$ obey the TDHF equation
under own self-consistent mean field of the event. According to the basic
postulate of the SMF approach, elements of the density matrix $\rho_{ji}^\lambda$
at the initial state have uncorrelated Gaussian distribution with the average values
$\overline{\rho_{ji}^\lambda}=n_j\delta_{ji}$ and the variances determined according to
\begin{equation}
\overline{\delta\rho_{ji}^\lambda\delta\rho_{i'j'}^\lambda}
= \frac{1}{2}\bigl[ n_j(1-n_i)+n_i(1-n_j) \bigr]\delta_{jj'}\delta_{ii'},
\label{Eq:SMF_dist}
\end{equation}
where $\delta\rho_{ji}^\lambda=\rho_{ji}^\lambda-n_j\delta_{ji}$ and
$n_j$ denotes the average occupation numbers of the single particle states.
Here and hereafter, the bar over quantities represents the ensemble average
over the stochastically generated events. At zero temperature the occupation
numbers are zero or one, while at finite temperatures they are specified according
to the Fermi-Dirac distribution. The distribution law (\ref{Eq:SMF_dist}) ensures
that an ensemble average of observables recovers the quantal expressions for
the mean and the variance at the initial state.

\begin{figure} [t]
\includegraphics[width=8cm]{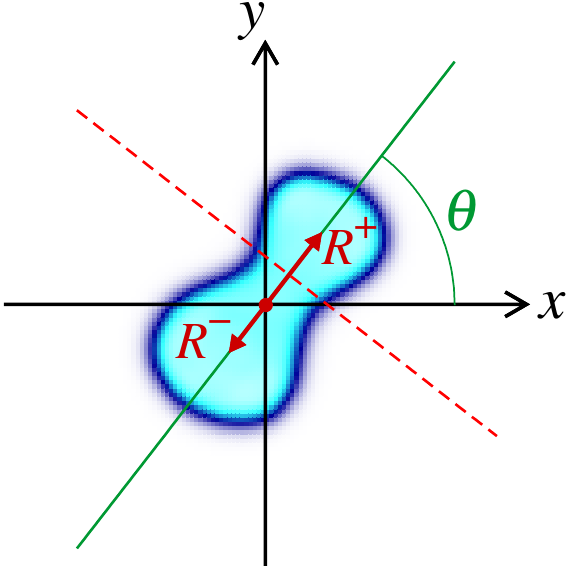}\vspace{-1mm}
\caption{
Density profile in the reaction plane at a certain instant in the $^{136}$Xe+$^{208}$Pb
reaction at \Ecm{526} with initial orbital angular momentum of $l=200\hbar$.
The beam direction is parallel to the $-x$ direction and the impact parameter vector
is parallel to $+y$ direction. The orientation angle of the dinuclear system is indicated
by $\theta$ ($=52.2^\circ$ at this instant). The red dot represents the center of mass
position of the system. The position vectors of projectile-like and target-like fragments
in the center-of-mass frame are indicated by $\bs{R}^+$ and $\bs{R}^-$, respectively,
where the relative distance at this instant is $R=|\bs{R}^+-\bs{R}^-|=13.7$\,fm. The
dashed line indicates the position of the window plane placed at the minimum density location.
}\vspace{-3mm}
\label{fig:schematic}
\end{figure}

In the special case, where colliding nuclei maintain a dinuclear structure (cf.\ 
Fig.~\ref{fig:schematic}, showing a typical density distribution in the $^{136}$Xe+$^{208}$Pb
reaction to be analyzed in Sec.~\ref{Sec:Results}), it is possible to analyze reaction
dynamics in terms of a few macroscopic variables, such as relative linear and angular
momenta, and mass and charge asymmetries of the dinuclear system. In this case,
the SMF approach gives rise to a set of coupled Langevin equations for the macroscopic
variables, which provides a quantal diffusion description of complex reaction dynamics
in terms of a few relevant macroscopic variables. With the quantal diffusion equations,
one can calculate not only the mean values of observables, which coincide with the TDHF
results, but also distributions of the observables. For details of the SMF approach
we refer readers to Refs.~\cite{Ayik(2008)1,Lacroix(2014)1,Ayik(2017),Ayik(2018)2}.
We also refer to recent applications of the SMF approach for the multinucleon transfer
mechanism in the dissipation heavy-ion collisions in Refs.~\cite{Ayik(2016),Ayik(2017),
Ayik(2018)2,Yilmaz(2018),Ayik(2019)1,Ayik(2019)2,Yilmaz(2020),
Sekizawa(2020)} and for kinetic energy fluctuations in spontaneous fission \cite{Tanimula(2018)}.

\subsection{Rate of change of the relative linear momentum}

In this section, let us recall basic equations that characterize the relative motion of colliding
nuclei. We define the relative distance, $\bs{R}(t)$, the reduced mass, $\mu(t)$,
and the relative linear momentum, $\bs{P}(t)$, in terms of the TDHF solutions with the
help of the window dynamics, see Fig.~\ref{fig:schematic}. Figure~\ref{fig:schematic}
illustrates the elongation axis (the solid line) and the window plane (the dashed line)
at a certain instant in the $^{136}$Xe+$^{208}$Pb reaction at \Ecm{526} with
the initial orbital angular momentum $l=200\hbar$. The orientation angle is indicated
by $\theta$ in the reaction plane. The elongation axis of the dinuclear system can
be determined by diagonalizing the mass quadruple tensor at any instant. The window
plane is perpendicular to the elongation axis and passes through the minimum density
location on the elongation axis. For description of the details of the window dynamics
we refer to Appendix~A in Ref.~\cite{Ayik(2018)2}.

In Fig.~\ref{fig:schematic}, the position vectors pointing the mean center-of-mass
position of projectile- and target-like fragments in the center-of-mass frame are
indicated by $\bs{R}^+$ and $\bs{R}^-$, respectively. In terms of the local density
$\rho_\lambda$ and the current density $\bs{j}_\lambda$ in the event $\lambda$,
the masses, the center-of-mass positions, and the linear momenta of the projectile-
and target-like fragments are, respectively, given by
\begin{eqnarray}
M_\lambda^\pm(t) &=& m\int d\bs{r}\, \Theta(\pm x')\,\rho_\lambda(\bs{r},t), \label{Eq:M_pm}\\
\bs{R}_\lambda^\pm(t) &=& m\int d\bs{r}\, \Theta(\pm x')\,\bs{r}\rho_\lambda(\bs{r},t)/M_\lambda^\pm(t), \\
\bs{P}_\lambda^\pm(t) &=& m\int d\bs{r}\, \Theta(\pm x')\,\bs{j}_\lambda(\bs{r},t),
\label{Eq:P_pm}
\end{eqnarray}
where $\bs{j}_\lambda$ denotes the current density in the event $\lambda$,
\begin{eqnarray}
\bs{j}_\lambda(\bs{r},t)
&=& \frac{\hbar}{2mi}\sum_{ij} \Bigl[
\phi_j^*(\bs{r},t;\lambda)\nabla\phi_i(\bs{r},t;\lambda) \nonumber\\[-2mm]
&&\hspace{12.5mm} -\phi_i(\bs{r},t;\lambda)\nabla\phi_j^*(\bs{r},t;\lambda)
\Bigr]\rho_{ji}^\lambda.
\end{eqnarray}
In Eqs.~(\ref{Eq:M_pm})--(\ref{Eq:P_pm}) we neglect the fluctuations in the window
geometry and specify the mean window position by a theta function $\Theta(\pm x')$,
where $x'(t)=[x-x_0(t)]\cos\theta(t)+[y-y_0(t)]\sin\theta(t)$ measures distance from
the window, $\theta(t)$ is the initially smaller angle between the elongation axis and the
beam direction, and $(x_0(t),y_0(t))$ is the position of the center of the window.

With the quantities introduced above, we can define the relative coordinate,
$\bs{R}_\lambda=\bs{R}_\lambda^+-\bs{R}_\lambda^-$, the reduced mass,
$\mu_\lambda=M_\lambda^+M_\lambda^-/(M_\lambda^++M_\lambda^-)$,
and the relative linear momentum,
\begin{eqnarray}
\bs{P}_\lambda
&=& \mu_\lambda\dot{\bs{R}}_\lambda
= \frac{M_\lambda^-\bs{P}_\lambda^+ - M_\lambda^+\bs{P}_\lambda^-}{M_\lambda^++M_\lambda^-} \nonumber\\[0.5mm]
&=&
\mu_\lambda[\dot{R}_\lambda\hat{\bs{e}}_R + R_\lambda\dot{\theta}_\lambda\hat{\bs{e}}_\theta].
\label{Eq:P_lambda}
\end{eqnarray}
Here, $\dot{\bs{R}}_\lambda=\dot{\bs{R}}_\lambda^+-\dot{\bs{R}}_\lambda^-$
denotes the relative velocity vector, where the velocities of the projectile- and target-like
fragments can be defined by $\dot{\bs{R}}_\lambda^\pm=\bs{P}_\lambda^\pm/M_\lambda^\pm$.
In the second line of Eq.~(\ref{Eq:P_lambda}), the relative velocity is decomposed into
the radial and tangential components with the unit vectors in respective directions,
\begin{eqnarray}
\hat{\bs{e}}_R &=& \cos\theta\,\hat{\bs{e}}_x + \sin\theta\,\hat{\bs{e}}_y, \\[1mm]
\hat{\bs{e}}_\theta &=& -\sin\theta\,\hat{\bs{e}}_x + \cos\theta\,\hat{\bs{e}}_y.
\end{eqnarray}
Neglecting the rate of change of the reduced mass, one finds the following
expression for the rate of change of the relative momentum:
\begin{eqnarray}
\frac{d\bs{P}_\lambda}{dt}
&=& \mu_\lambda \bigl[ (\ddot{R}_\lambda - R_\lambda\dot{\theta}_\lambda^2)\hat{\bs{e}}_R
+ (R_\lambda\ddot{\theta}_\lambda + 2\dot{R}_\lambda\dot{\theta}_\lambda)\hat{\bs{e}}_\theta \bigr] \nonumber\\[0.5mm]
&=&
\biggl( \frac{dK_\lambda}{dt} - \frac{L_\lambda^2}{\mu_\lambda R_\lambda^3} \biggr)\hat{\bs{e}}_R
+\biggl( \frac{1}{R_\lambda}\frac{dL_\lambda}{dt} \biggr)\hat{\bs{e}}_\theta,
\label{Eq:dPdt}
\end{eqnarray}
where we have introduced the radial component of the relative linear momentum
$K_\lambda$ and the relative orbital angular momentum $L_\lambda$ defined as
\begin{eqnarray}
K_\lambda &\equiv& \hat{\bs{e}}_R\bs{\cdot}\bs{P}_\lambda, \\[1mm]
L_\lambda &\equiv& \mu_\lambda R_\lambda^2\dot{\theta}.
\end{eqnarray}
The first and the second terms of Eq.~(\ref{Eq:dPdt}) denote
the rate of changes of the radial and the tangential components,
respectively.

\subsection{Stochastic equations for the relative momenta}

In the SMF approach, we can express the rate of change of the
projectile- and target-like fragments in an event $\lambda$ as
\begin{equation}
\frac{d\bs{P}_\lambda^\pm}{dt}
= \pm m\int d\bs{r} \delta(x')\dot{x}'\bs{j}_\lambda(\bs{r},t)
+ m\int d\bs{r} \Theta(\pm x')\frac{\partial\bs{j}_\lambda(\bs{r},t)}{\partial t}.
\end{equation}
Employing the TDHF equation for the single-particle orbitals in the event
$\lambda$, it is possible to write down the rate of change of the radial
and the tangential components of the linear momentum of the fragments
in the following form:
\begin{eqnarray}
\hat{\bs{e}}_\alpha\bs{\cdot}\frac{d\bs{P}_\lambda^\pm}{dt}
&=& \pm\int d\bs{r}\, \delta(x')\dot{x}'m\,\hat{\bs{e}}_\alpha\bs{\cdot}\bs{j}_\lambda(\bs{r},t) \nonumber\\
&&-\int d\bs{r}\, \Theta(\pm x')\nabla\bs{\cdot}\sum_{ij}\Bigl( \bs{A}_{ji}^\alpha - \bs{B}_{ji}^\alpha \Bigr)\rho_{ji}^\lambda \nonumber\\
&&+\;\mbox{[Potential terms]}
\label{Eq:ROC_P}
\end{eqnarray}
with
\begin{eqnarray}
\bs{A}_{ji}^\alpha &=& \frac{\hbar^2}{4m}\Bigl[ \phi_i(\bs{r},t;\lambda) \nabla\bigl( \hat{\bs{e}}_\alpha\bs{\cdot}\nabla\phi_j^*(\bs{r},t;\lambda) \bigr) + \mbox{c.c.} \Bigr], \label{Eq:def_Amat}\\
\bs{B}_{ji}^\alpha &=& \frac{\hbar^2}{4m}\Bigl[ \bigl( \nabla\phi_i(\bs{r},t;\lambda) \label{Eq:def_Bmat}\bigr)
\hat{\bs{e}}_\alpha\bs{\cdot}\nabla\phi_j^*(\bs{r},t;\lambda) + \mbox{c.c.} \Bigr],
\end{eqnarray}
where $\hat{\bs{e}}_\alpha$ indicates the unit vector in the radial ($\alpha=R$)
or the tangential ($\alpha=\theta$) direction. The `[Potential terms]' in Eq.~(\ref{Eq:ROC_P})
represents terms associated with mean-field potentials other than the kinetic term
in the single-particle Hamiltonian in the event $\lambda$.

From Eq.~(\ref{Eq:ROC_P}), we can derive a Langevin equation
for the rate of change of the relative linear momentum,
\begin{equation}
\frac{d\bs{P}_\lambda}{dt} = \int d\bs{r}\,g(x')\,\dot{x}'m\bs{j}_\lambda(\bs{r},t)
+ \mbox{[Potential terms]} + \bs{f}^\lambda(t),
\label{Eq:ROC_Prel}
\end{equation}
where the first and the second terms represent the forces arising from the motion
of the window plane and the potential terms, respectively. The quantity $\bs{f}^\lambda(t)$
is the fluctuating dynamical force due to nucleon exchange between projectile- and
target-like fragments. Its radial ($\alpha=R$) and tangential ($\alpha=\theta$)
components are given by
\begin{eqnarray}
f_\alpha^\lambda(t) = \sum_{ij}Y_{ji}^\alpha(t)\rho_{ji}^\lambda,
\end{eqnarray}
where we have introduced a shorthand notation,
\begin{equation}
Y_{ji}^\alpha(t) \equiv 
\int d\bs{r}\,g(x')\,\hat{\bs{e}}_R\bs{\cdot}\Bigl( \bs{A}_{ji}^\alpha(t) - \bs{B}_{ji}^\alpha(t) \Bigr).
\label{Eq:Y}
\end{equation}
In obtaining this result we employed a partial integration in Eq.~(\ref{Eq:ROC_P})
and used the following relations:
\begin{eqnarray}
\frac{\partial}{\partial x}\Theta(x') &=& \delta(x')\cos\theta, \\[1mm]
\frac{\partial}{\partial y}\Theta(x') &=& \delta(x')\sin\theta.
\end{eqnarray}
In Eq.~(\ref{Eq:ROC_Prel}) the delta function has been replaced with
a smoothing function $\delta(x')\rightarrow g(x')$ expressed as a Gaussian,
\begin{equation}
g(x') = \frac{1}{\sqrt{2\pi}\kappa}\exp\Bigl[ -\frac{1}{2}\Bigl(\frac{x'}{\kappa}\Bigr)^2 \Bigr],
\label{Eq:g}
\end{equation}
with a dispersion $\kappa=1.0$\,fm which is on the same order as the lattice
spacing in the numerical calculations. By projecting Eq.~(\ref{Eq:ROC_Prel})
along the radial and the tangential directions, together with Eq.~(\ref{Eq:dPdt}),
we obtain two coupled Langevin equations for the radial and angular momenta
\cite{Gardiner(1991),Weiss(1999)}:
\begin{eqnarray}
\frac{dK_\lambda}{dt}-\frac{L_\lambda^2}{\mu_\lambda R_\lambda^3}
&=& \int d\bs{r}\, g(x')\,\dot{x}'m\,\hat{\bs{e}}_R\bs{\cdot}\bs{j}_\lambda(\bs{r},t) + f_R^\lambda(t) \nonumber\\[0.5mm]
&&+ \mbox{[Potential terms]},
\label{Eq:dKdt}\\[2.5mm]
\frac{1}{R_\lambda}\frac{dL_\lambda}{dt}
&=& \int d\bs{r}\, g(x')\,\dot{x}'m\,\hat{\bs{e}}_\theta\bs{\cdot}\bs{j}_\lambda(\bs{r},t) + f_\theta^\lambda(t) \nonumber\\[0.5mm]
&&+ \mbox{[Potential terms]}.
\label{Eq:dLdt}
\end{eqnarray}
In the right-hand side of these expressions, the first and the third terms
represent the force due to the motion of the window plane and the conservative
force due to nuclear and electrical potential energies, respectively. The fluctuating
forces, $f_R^\lambda(t)$ and $f_\theta^\lambda(t)$, represent the dynamical
forces arising from nucleon exchange between projectile- and target-like fragments.
These dynamical forces provide the dominant mechanism for the dissipation
and fluctuations of the relative momentum in damped collisions of heavy ions,
such as deep-inelastic and quasifission processes.

The ensemble average of these equations of motion are equivalent to the TDHF
description for the radial and angular components of the relative linear momentum.
Consequently, we use the mean values of TKE and the orbital angular momentum
obtained from the TDHF approach. We employ the Langevin equations, Eqs.~(\ref{Eq:dKdt})
and (\ref{Eq:dLdt}), for describing fluctuations around their mean values. There
are two different sources for fluctuations of the dynamical forces $f_R^\lambda(t)$
and $f_\theta^\lambda(t)$ induced by
nucleon exchange: (\textit{i}) fluctuations due to different set of wave functions
in each event $\lambda$, and (\textit{ii}) fluctuations introduced by the stochastic
part $\delta\rho_{ji}^\lambda$ of the density matrix at the initial sate. The former
part of fluctuations can be approximately described in terms of the fluctuating
components of the radial and angular momentum as $f_R^\lambda(t) \rightarrow
f_R^{\rm diss}(K_\lambda)$ and $f_\theta^\lambda(t) \rightarrow f_\theta
^{\rm diss}(L_\lambda)$. Here, $f_R^{\rm diss}(K_\lambda)$ and $f_\theta
^{\rm diss}(L_\lambda)$ are the mean values of the radial and tangential
components of the dissipative part of the dynamical forces expressed in terms
of fluctuating radial and angular momenta, respectively. We assume that
the amplitude of the fluctuations are sufficiently small, so that we can linearize
the Langevin equations, Eq.~(\ref{Eq:dKdt}) and (\ref{Eq:dLdt}), around
the mean values to give
\begin{eqnarray}
\frac{\partial}{\partial t}\delta K_\lambda - \frac{2L}{\mu R^3}\delta L_\lambda
&=& \Bigl(\frac{\partial f_R^{\rm diss}}{\partial K}\Bigr)\delta K_\lambda + \delta f_R^\lambda,
\label{Eq:ddKdt}\\[2mm]
\frac{\partial}{\partial t}\delta L_\lambda
&=& \Bigl(\frac{\partial f_\theta^{\rm diss}}{\partial L}\Bigr)\delta L_\lambda + R\,\delta f_\theta^\lambda,
\label{Eq:ddLdt}
\end{eqnarray}
where $\delta K_\lambda=K_\lambda-\overline{K_\lambda}$ and
$\delta L_\lambda=L_\lambda-\overline{L_\lambda}$ are the fluctuating
components of the radial and angular momenta, respectively. The fluctuating forces
originating from the potential energy terms are expected to have a small effect
on the fluctuations of the relative momentum. In these expressions, we neglect
these forces as well as the force due to the motion of the window given
by the first terms in the right-hand side of Eqs.~(\ref{Eq:dKdt}) and (\ref{Eq:dLdt}).
Also, we neglect the fluctuations in the reduced mass and the relative distance
between the centers of the fragments. The quantities, $\mu(t)$, $R(t)$, and $L(t)$,
are, respectively, the mean values of the reduced mass, the relative distance, and
the relative orbital angular momentum of the colliding system, which are determined
by the TDHF equation. The derivatives of dissipative forces on the right-hand side
of Eqs.~(\ref{Eq:ddKdt}) and (\ref{Eq:ddLdt}) are related to the reduced
radial and tangential friction coefficients:
\begin{eqnarray}
\frac{\partial f_R^{\rm diss}}{\partial K} &=& -\gamma_R(t), \\
\frac{\partial f_\theta^{\rm diss}}{\partial L} &=& -\gamma_\theta(t).
\end{eqnarray}
An analysis of the radial friction force and the reduced friction coefficients
are presented in Appendix~\ref{App:gamma_R}.

Multiplying both sides of Eqs.~(\ref{Eq:ddKdt}) and (\ref{Eq:ddLdt}) by $\delta K_\lambda$
and $\delta L_\lambda$, respectively, and taking the ensemble average, we obtain a set of
coupled differential equations for the variances \cite{Merchant(1982),Schroeder(1981),Risken(1996)},
\begin{eqnarray}
\frac{d\sigma_{KK}^2}{dt} - \frac{4L}{\mu R^3}\sigma_{KL}^2
&=& -2\gamma_R\sigma_{KK}^2 + 2D_{KK}, \label{Eq:PDE1}\\[0.5mm]
\frac{d\sigma_{LL}^2}{dt}
&=& -2\gamma_\theta\sigma_{LL}^2 + 2R^2D_{LL}, \\[0.5mm]
\frac{d\sigma_{KL}^2}{dt} - \frac{2L}{\mu R^3}\sigma_{LL}^2
&=& -2(\gamma_R+\gamma_\theta)\sigma_{KL}^2 + R(D_{KL}+D_{LK}),\nonumber\\[-2mm]
\end{eqnarray}
where the variances are defined as $\sigma_{KK}^2(t)=\overline{\delta K_\lambda(t)
\delta K_\lambda(t)}$, $\sigma_{LL}^2(t)=\overline{\delta L_\lambda(t)\delta L_\lambda(t)}$,
and $\sigma_{KL}^2(t)=\overline{\delta K_\lambda(t)\delta L_\lambda(t)}$.
Here, $D_{\alpha\beta}(t)$ denotes the momentum diffusion coefficients
($\alpha,\beta=K,L)$\footnote{Note that we use the same notation
($\alpha,\beta$) to indicate the radial and tangential directions ($R,\theta$) and
the radial and angular momenta ($K,L$).}, which are expressed on a microscopic
basis in terms of single-particle orbitals within the TDHF approach.

\section{Total kinetic energy distribution \\within the SMF approach}\label{Sec:TKE}

\subsection{Momentum diffusion coefficients}\label{Sec:DiffusionCoef}

The momentum diffusion coefficients for the radial and angular momenta
are defined as the time integral over the history of the autocorrelation
functions of the stochastic forces,
\begin{equation}
D_{\alpha\beta}(t) = \int_0^t dt'\,\overline{\delta f_\alpha^\lambda(t)\delta f_\beta^\lambda(t')}.
\end{equation}
The stochastic parts of the radial and tangential forces are given by,
\begin{equation}
\delta f_\alpha^\lambda(t)
= \sum_{ij}Y_{ji}^\alpha(t)\delta\rho_{ji}^\lambda.
\label{Eq:delta_f}
\end{equation}
Using the basic postulate of the SMF approach, we can analytically take
the ensemble average, and the correlation functions of the random force
on radial and tangential directions read
\begin{eqnarray}
\overline{\delta f_\alpha^\lambda(t)\delta f_\beta^\lambda(t')}
&=& {\rm Re}\Biggl[\; \sum_{p\in{\rm P},h\in{\rm T}}Y_{hp}^\alpha(t)Y_{hp}^{\beta*}(t') \nonumber\\[-1mm]
&&\hspace{4mm}+ \sum_{p\in{\rm T},h\in{\rm P}}Y_{hp}^\alpha(t)Y_{hp}^{\beta*}(t') \;\Biggr].
\label{Eq:bar_dfdf}
\end{eqnarray}
In this expression, the summations in the first term run over the particle states
originating from the projectile $p\in{\rm P}$ and the hole states originating from the
target $h\in{\rm T}$, while in the second term the summations run in the opposite way.
By adding and subtracting the hole-hole terms, the first term in this expression
can be written as,
\begin{eqnarray}
\sum_{p\in{\rm P},h\in{\rm T}}Y_{hp}^\alpha(t)Y_{hp}^{\beta*}(t')
&=& \sum_{h\in{\rm T},a\in{\rm P}}Y_{ha}^\alpha(t)Y_{ha}^{\beta*}(t') \nonumber\\
&-& \sum_{h\in{\rm T},h'\in{\rm P}}Y_{hh'}^\alpha(t)Y_{hh'}^{\beta*}(t').
\end{eqnarray}
In the first term, the summation $a\in{\rm P}$ runs over the complete set of states
originating from the projectile. We introduce a similar subtraction in the second
term of Eq.~(\ref{Eq:bar_dfdf}). As shown in Appendix~\ref{App:A}, using the closure relation in a diabatic
approximation of the TDHF orbitals, it is possible to eliminate the complete set
of the projectile (target) states in the first (second) term. As a result, the radial,
the tangential and the mixed diffusion coefficients are given by the following compact expression:
\begin{eqnarray}
D_{\alpha\beta}(t)
&=& \int_0^t\hspace{-0.5mm}d\tau \int d\bs{r}\,\tilde{g}(x')\,\Bigl[\, G_{\rm T}(\tau)J_{\alpha\beta}^{\rm T}(\bs{r},\bar{t}) \nonumber\\
&&\hspace{26mm}+\;G_{\rm P}(\tau)J_{\alpha\beta}^{\rm P}(\bs{r},\bar{t}) \,\Bigr] \nonumber\\[0mm]
&-& \int_0^t\hspace{-0.5mm}d\tau\,{\rm Re}\Biggl[ \sum_{h\in{\rm T},h'\in{\rm P}}Y_{hh'}^\alpha(t)Y_{hh'}^{\beta*}(t-\tau) \nonumber\\[-2mm]
&&\hspace{12mm}
+\hspace{-0.5mm}\sum_{h\in{\rm P},h'\in{\rm T}}Y_{hh'}^\alpha(t)Y_{hh'}^{\beta*}(t-\tau) \Biggr].
\label{Eq:D_alphabeta}
\end{eqnarray}
In the first line, the quantity $J_{\alpha\beta}^{\rm T}(\bs{r},\bar{t})$ is given by
\begin{eqnarray}
J_{\alpha\beta}^{\rm T}(\bs{r},\bar{t})
&=& \frac{\hbar}{m}\sum_{h\in{\rm T}}\bigl[mu_\alpha^h(\bs{r},\bar{t})\bigr]\bigl[mu_\beta^h(\bs{r},\bar{t})\bigr] \nonumber\\
&&\times\; \Bigl| {\rm Im}\bigl[\phi_h^*(\bs{r},\bar{t})\hat{\bs{e}}_R\bs{\cdot}\nabla\phi_h(\bs{r},\bar{t})\bigr] \Bigr|,
\label{Eq:J_T}
\end{eqnarray}
where $\bar{t}=(t+t')/2=t-\tau/2$. This expression represents the magnitude
of the nucleon flux that carries the product of the momentum components
$mu_\alpha^h(\bs{r},\bar{t})$ and $mu_\beta^h(\bs{r},\bar{t})$ from the
target-like fragment in the perpendicular ($\alpha,\beta=R$) and tangential ($\alpha,
\beta=\theta$) directions to the window plane. The quantity $J_{\alpha\beta}
^{\rm P}(\bs{r},t-\tau/2)$ is given by a similar expression and it represents
the magnitude of the nucleon flux from the projectile-like fragment. The radial
and tangential components of the nucleon flow velocities are determined by
\begin{eqnarray}
u_\alpha^h(\bs{r},\bar{t}) 
&=& \frac{\hbar}{m}\frac{{\rm Im}\bigl[ \phi_h^{\alpha*}(\bs{r},\bar{t})\hat{\bs{e}}_\alpha\bs{\cdot}\nabla\phi_h^\alpha(\bs{r},\bar{t}) \bigr]}{\bigl|\phi_h(\bs{r},\bar{t})\bigr|^2}.
\end{eqnarray}

We observe that there is a close analogy between the quantal expression of the
diffusion coefficients and the classical ones in a random walk problem. The first
term in the quantal expression (\ref{Eq:D_alphabeta}) gives the sum of the
nucleon flux across the window from the target-like to the projectile-like fragments
and vise versa, which is integrated over the memory. Each nucleon transfer across
the window in both directions carries the product of the momentum components
which increases the rate of change of the momentum dispersion. This is analogous
to the random walk problem, in which the diffusion coefficient is given by the sum
of the rate for forward and backward steps. The second term in the quantal expression
(\ref{Eq:D_alphabeta}) stands for the Pauli blocking effects in the nucleon transfer
mechanism, which does not have a classical counterpart. The quantities in the Pauli
blocking factors are determined by hole-hole elements of the matrices $Y_{hh'}^\alpha(t)$
and $Y_{hh'}^{\beta*}(t)$ which are defined in Eq.~(\ref{Eq:Y}) with
Eqs.~(\ref{Eq:def_Amat}) and (\ref{Eq:def_Bmat}).

\subsection{Total kinetic energy distribution}


It is possible to determine the joint probability distribution function of the radial
linear momentum $K$ and the orbital angular momentum $L$ for each initial
orbital angular momentum $l$, $P_l(K,L)$, employing the coupled Langevin
equations, Eqs.~(\ref{Eq:ddKdt}) and (\ref{Eq:ddLdt}). It is well known that
these coupled Langevin equations are equivalent to the Fokker-Planck description
for the joint probability distribution $P_l(K,L)$ \cite{Gottfried(1966)}. When the
radial and tangential friction forces have linear dependence on the radial and the
angular momenta, the solution of the joint probability distribution can be
expressed as a correlated Gaussian function:
\begin{equation}
P_l(K,L) = \frac{\exp[-C_l(K,L)]}{2\pi\sigma_{KK}(l)\sigma_{LL}(l)\sqrt{1-\eta_l^2}},
\label{Eq:P_l}
\end{equation}
where
\begin{eqnarray}
C_l(K,L) &=& \frac{1}{2(1-\eta_l^2)}
\Biggl[ \biggl(\frac{K-K_l}{\sigma_{KK}(l)}\biggr)^2 + \biggl(\frac{L-L_l}{\sigma_{LL}(l)}\biggr)^2 \nonumber\\[1mm]
&&\hspace{13mm}-2\eta_l\biggl(\frac{K-K_l}{\sigma_{KK}(l)}\biggr)\biggl(\frac{L-L_l}{\sigma_{LL}(l)}\biggr) 
\Biggr].
\end{eqnarray}
Here, the correlation factor is defined as $\eta_l=\sigma_{KL}^2(l)/\sigma_{KK}(l)
\sigma_{LL}(l)$. $K_l\equiv\overline{K_\lambda(l)}$ and $L_l\equiv\overline{L_\lambda(l)}$
denote the mean values of the radial and the angular momenta for each value of the initial orbital
angular momentum $l$, respectively, which are determined by solving the TDHF equation.

The mean values of the radial and angular momenta, $K_l$ and $L_l$, are
obtained by solving the TDHF equation. In practice, we follow the reaction
dynamics up to a certain instant, say $t=t_{\rm f}$, at which binary products
are well separated spatially. Denoting the relative distance at this instant as
$R_{\rm f}=R(t_{\rm f})$, the asymptotic value of TKE of the outgoing
fragments is given by $E_{\rm kin}^\infty(K,L)=K^2/2\mu+L^2/2\mu R_{\rm f}^2
+Z_{1}Z_{2}e^2/R_{\rm f}$. For a given initial angular momentum $l$,
we define the TKE distribution $G_l(E)$ as
\begin{eqnarray}
G_l(E) &=& \int dKdL\;\delta\bigl( E-E_{\rm kin}^\infty(K,L) \bigr) P_l(K,L).
\label{Eq:G_l(K,L)}
\end{eqnarray}
Note that $K$ and $L$ in the above expression correspond to the radial and the
angular momenta at the instant $t=t_{\rm f}$, respectively, and $E$ stands here
for the asymptotic TKE. It is to mention that $\mu$ and $Z_{1,2}$ are, in general,
$l$ dependent quantities, and the fluctuations in the mass and charge asymmetries
may affect the TKE fluctuations. However, we neglect the effects of mass and charge
fluctuations on the TKE distribution and retain the mean values of the mass and
charge asymmetry for each angular momentum.

In practice, the mixed diffusion coefficients, $D_{KL}(t)$ and $D_{LK}(t)$,
are expected to be much smaller than the radial and the angular momentum diffusion
coefficients, $D_{KK}(t)$ and $D_{LL}(t)$. Hence, in the present work, we
neglect the mixed dispersion term $\sigma_{KL}(t)$ in Eq.~(\ref{Eq:PDE1})
and the coupling between the radial and angular momenta. In such a case, the
expression can be greatly simplified by taking the asymptotic limit,
$R\rightarrow\infty$, leading to
\begin{equation}
G_l(E) = \int dK^\infty\;\delta\bigl( E-E_{\rm kin}^\infty(K^\infty) \bigr) P_l(K^\infty),
\label{Eq:G_l(K)}
\end{equation}
where $E_{\rm kin}^\infty(K)=K^2/2\mu$ and $P_l(K)$ is the probability
distribution of the radial momentum. Notice that by taking the limit $R\rightarrow\infty$
the centrifugal part of the kinetic energy and the Coulomb energy entirely transformed
into the radial TKE, and $K^\infty$ in Eq.~(\ref{Eq:G_l(K)}) corresponds
to the asymptotic value of the radial momentum for $R\rightarrow\infty$.
After taking the integral over the angular momentum variable, the asymptotic
radial momentum distribution becomes a simple Gaussian,
\begin{equation}
P_l(K) = \frac{1}{\sqrt{2\pi}\sigma_{KK}(l)}\exp\biggl[ -\frac{1}{2}\biggl(\frac{K-K_l^\infty}{\sigma_{KK}(l)}\biggr)^2 \biggr],
\end{equation}
where the mean value of the asymptotic radial momentum is related to the mean
asymptotic TKE from TDHF, $E_{\rm kin}^\infty(l)$, by $K_l^\infty=\bigl(2\mu
E_{\rm kin}^\infty(l)\bigr)^{1/2}$. After a trivial integration, we obtain the
asymptotic TKE distribution,
\begin{eqnarray}
G_l(E) &=& \frac{1}{\sqrt{8\pi E}\,\widetilde{\sigma}_{KK}(l)}
\exp\Biggl[ -\frac{1}{2}\Biggl(\frac{\sqrt{E}-\sqrt{E_{\rm kin}^\infty(l)}}{\widetilde{\sigma}_{KK}(l)}\Biggr)^2 \Biggr], \nonumber\\
\label{Eq:TKE_dist}
\end{eqnarray}
where $\widetilde{\sigma}_{KK}(l)\equiv\sigma_{KK}(l)/\sqrt{2\mu}$.
To obtain the radial dispersion $\sigma_{KK}(t)$, we solve the quantal diffusion
equation for the radial component,
\begin{equation}
\frac{d\sigma_{KK}^2}{dt} = -2\gamma_R\sigma_{KK}^2 + 2D_{KK}.
\label{Eq:PDE_sigma2_KK}
\end{equation}
We note that the unit of the TKE distribution $G_l$ is MeV$^{-1}$, hence
the fraction of events with final TKE in the energy range $\Delta E$ in MeV
is given by $G_l\Delta E$.

\begin{table}[t]
\caption{
A list of numerical results of the TDHF calculations for the $^{136}$Xe+$^{208}$Pb
reaction at \Ecm{526}. From left to right columns, it shows: the initial orbital angular
momentum, $l$, in $\hbar$, the final average relative orbital angular momentum, $L_{\rm f}$,
in $\hbar$, neutron and proton numbers of projectile-like (target-like) fragment, $N_1$
and $Z_1$ ($N_2$ and $Z_2$), mean total kinetic energy loss (TKEL) in MeV, contact
time, $t_{\rm contact}$, in fm/$c$, scattering angles in center-of-mass frame, $\theta_{\rm c.m.}$,
and those in laboratory frame for projectile-like (target-like) fragment, $\vartheta_1^{\rm lab}$
($\vartheta_2^{\rm lab}$), in degrees. The contact time is defined as duration in which
the minimum density between two colliding nuclei exceeds half the saturation density,
$\rho_{\rm sat}/2=0.08$\,fm$^{-3}$.
}\vspace{-1mm}
\label{Table:TDHF}
\begin{center}
\begin{tabular*}{\columnwidth}{@{\extracolsep{\fill}}ccccccccccc}
\hline\hline
$l$	&$L_{\rm f}$&	$N_1$&	$Z_1$&	$N_2$&	$Z_2$&	TKEL&	$t_{\rm contact}$&	$\theta_{\rm c.m.}$&	$\vartheta_1^{\rm lab}$&	$\vartheta_2^{\rm lab}$ \\[-0.6mm]
($\hbar$) & ($\hbar$) & {} & {} & {} & {} & (MeV) & (fm/$c$) & (deg) & (deg) & (deg) \\
\hline
0	&0&  	82.3&	55.1&	125.0&	80.9&	173.1&	661.4&     	180.0&     	180.0&	0.0 \\
50	&39&	82.4&	55.1&	124.9&	80.8&	177.0&	646.0&     	150.1&     	96.4&	13.5 \\
100	&78&	82.0&	54.5&	125.4&	81.5&	175.6&	611.8&     	123.1&     	72.9&	25.4 \\
110	&85&	81.9&	54.3&	125.5&	81.6&	175.1&	593.2&     	118.2&     	69.5&	27.5 \\
120	&95&	81.8&	54.3&	125.6&	81.7&	175.0&	591.4&     	113.3&     	66.2&	29.6 \\
130	&104&	81.7&	54.3&	125.7&	81.7&	175.3&	586.6&     	108.6&     	63.0&	31.6 \\
140	&114&	81.9&	54.4&	125.5&	81.5&	175.0&	571.0&     	104.0&     	60.0&	33.6 \\
150	&124&	82.4&	54.7&	125.0&	81.2&	174.3&	555.2&     	99.7&          	57.1&	35.6 \\
160	&134&	83.1&	55.2&	124.3&	80.7&	171.9&	538.6&     	95.9&          	54.6&	37.4 \\
170	&142&	83.7&	55.6&	123.7&	80.3&	169.4&	529.4&     	92.5&          	52.5&	39.0 \\
180	&149&	83.9&	55.8&	123.5&	80.2&	168.5&	517.6&     	89.7&          	50.7&	40.3 \\
190	&158&	83.6&	55.6&	123.9&	80.3&	167.4&	474.0&     	87.1&          	49.3&	41.3 \\
200	&166&	83.0&	55.4&	124.5&	80.6&	166.0&	462.4&     	85.2&          	48.4&	42.0 \\
210	&173&	82.7&	55.2&	124.9&	80.7&	161.0&	440.2&     	84.1&          	47.9&	42.6 \\
220	&179&	82.3&	55.0&	125.3&	80.9&	154.9&	409.4&     	83.3&          	47.8&	43.1 \\
230	&185&	81.8&	54.8&	125.8&	81.2&	149.1&	378.4&     	82.5&          	47.6&	43.5 \\
240	&194&	81.5&	54.6&	126.2&	81.3&	140.7&	343.6&     	81.7&          	47.4&	44.1 \\
250	&203&	81.3&	54.5&	126.4&	81.5&	129.8&	303.2&     	81.0&          	47.4&	44.8 \\
260	&214&	81.2&	54.4&	126.6&	81.5&	117.1&	257.2&     	80.5&          	47.5&	45.5 \\
270	&225&	81.2&	54.4&	126.6&	81.5&	103.5&	226.0&     	80.0&          	47.5&	46.2 \\
280	&240&	81.4&	54.5&	126.5&	81.5&	86.0&	192.4&     	79.5&          	47.6&	47.2 \\
290	&258&	81.5&	54.5&	126.4&	81.5&	66.8&	146.8&     	79.0&          	47.8&	48.2 \\
300	&277&	81.5&	54.4&	126.4&	81.6&	48.9&	100.8&     	78.5&          	47.9&	49.0 \\
310	&295&	81.6&	54.3&	126.4&	81.7&	32.7&	44.2&          	78.1&          	48.0&	49.8 \\
320	&311&	81.7&	54.2&	126.2&	81.8&	16.9&	0.0&           	77.9&          	48.2&	50.4 \\
330	&325&	81.9&	54.1&	126.1&	81.9&	8.0& 	0.0&           	77.4&          	48.1&	51.0 \\
340	&337&	81.9&	54.0&	126.1&	82.0&	4.7& 	0.0&           	76.4&          	47.5&	51.6 \\
350	&347&	81.9&	54.0&	126.1&	82.0&	3.3& 	0.0&           	75.2&          	46.7&	52.3 \\

\hline\hline
\end{tabular*}\vspace{-6mm}
\end{center}
\end{table}

\section{Results for Xe+Pb Collisions}\label{Sec:Results}

In this section, as the first application of the proposed formalism given
in the preceding sections, we present calculations of the TKE distribution
for the $^{136}$Xe+$^{208}$Pb reaction at \Ecm{526}, for which
extensive experimental data reported by Kozulin \textit{et al.} \cite{Kozulin(2012)}
are available. TDHF calculations were carried out for a range of initial
orbital angular momenta $l$. The results of TDHF calculations for a set
of observables in the $^{136}$Xe+$^{208}$Pb reaction at \Ecm{526}
are presented in Table~\ref{Table:TDHF}. We mention here that for
the $^{136}$Xe+$^{208}$Pb system the average numbers of
transferred nucleons are small, reflecting a small charge asymmetry
and possible shell effects in the reactants. Nucleons are, however,
actively exchanged during the collision, which is the source of dissipation
and fluctuations of observables, such as mass, charge, TKE, and scattering
angles, in low-energy heavy-ion reactions. For this reaction, mean TKEL
reaches around 175\,MeV for small angular momenta, while contact time
is rather short ($\lesssim 2$\,zs). Because of the short contact time the
composite system does not rotate much in the reaction plane. We note
that fragments are emitted outside the experimental angular coverage
(25$^\circ$--70$^\circ$ in the laboratory frame) in events below
$l<100\hbar$ in TDHF calculations.

In order to evaluate the TKE distribution, we have further extended own
three-dimensional parallel TDHF code, which was applied for various systems
\cite{KS_KY_MNT,KS_KY_PNP,Bidyut(2015),KS_KY_Ni-U,KS_U-Sn,Bidyut(2017),
KS_GEMINI,Williams(2018)} and was recently incorporated with the SMF approach
\cite{Sekizawa(2020)}. For the computational details we refer readers to our recent
article, Ref.~\cite{Sekizawa(2020)}. To obtain the TKE distribution (\ref{Eq:TKE_dist}),
we need to evaluate the asymptotic value of the radial momentum dispersion
$\sigma_{KK}(l)$ for each value of the initial angular momentum $l$ by solving
Eq.~(\ref{Eq:PDE_sigma2_KK}). The radial momentum diffusion coefficient
$D_{KK}(t)$ is directly computed from occupied single-particle orbitals within the
TDHF approach with the quantal expression given in Eq.~(\ref{Eq:D_alphabeta}).
On the other hand, it is not trivial how to determine the radial friction coefficient directory
from TDHF. Nevertheless, using the analogy to the random walk problem, we have extracted
from TDHF an approximate expression for the radial friction force and the radial friction
coefficients. Details of this analysis are given in Appendix~\ref{App:gamma_R}.

\begin{figure} [t]
\includegraphics[width=8.6cm]{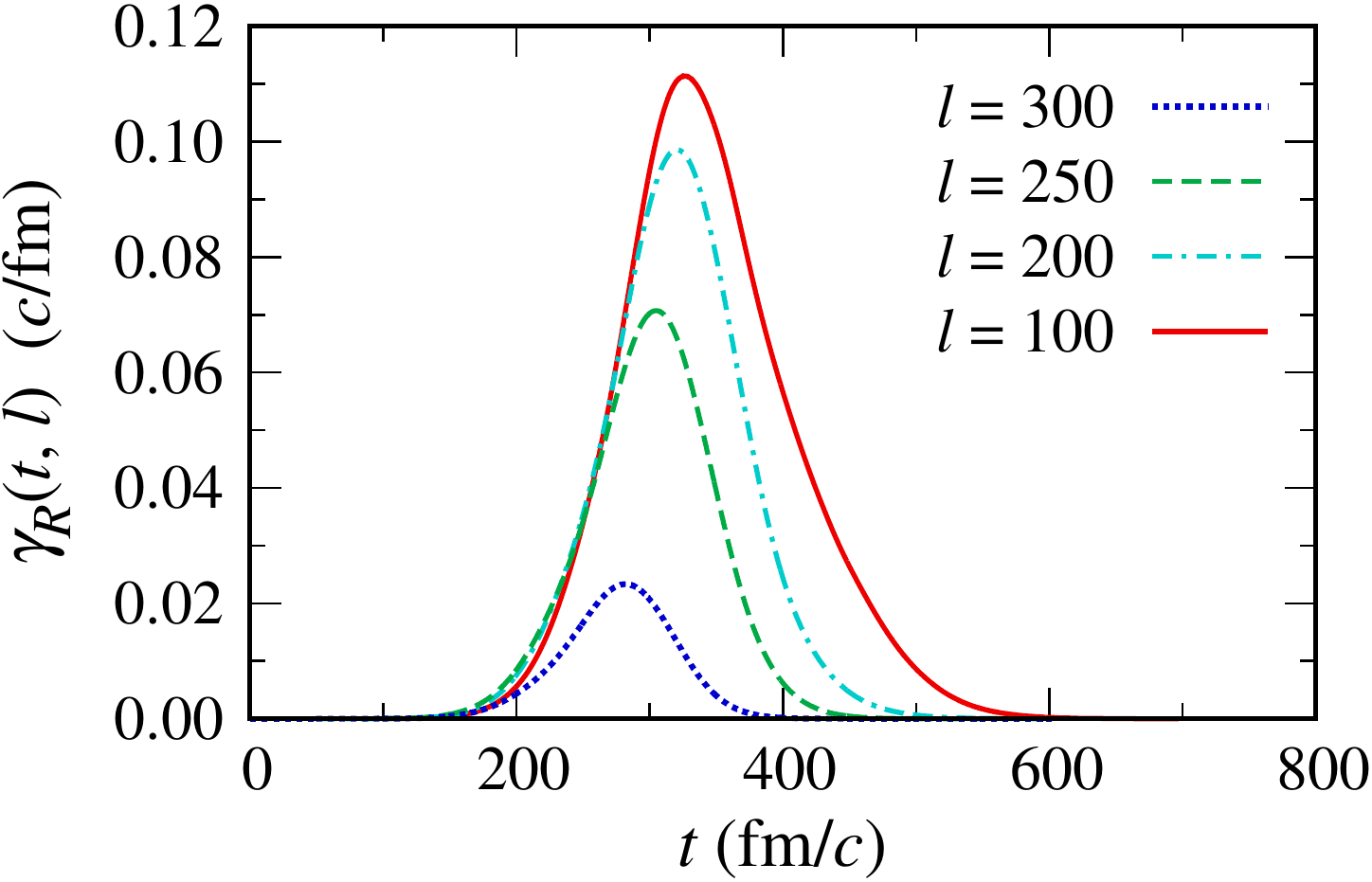}\vspace{-1mm}
\caption{
Reduced radial friction coefficients $\gamma_R$ in the $^{136}$Xe+$^{208}$Pb reaction
at \Ecm{526} with initial orbital angular momenta of $l$\,$=$\,100, 200, 250, and 300
(in units of $\hbar$) are shown as functions of time.
}\vspace{-3mm}
\label{fig:gamma_R}
\end{figure}

\begin{figure} [b]
\includegraphics[width=8.6cm]{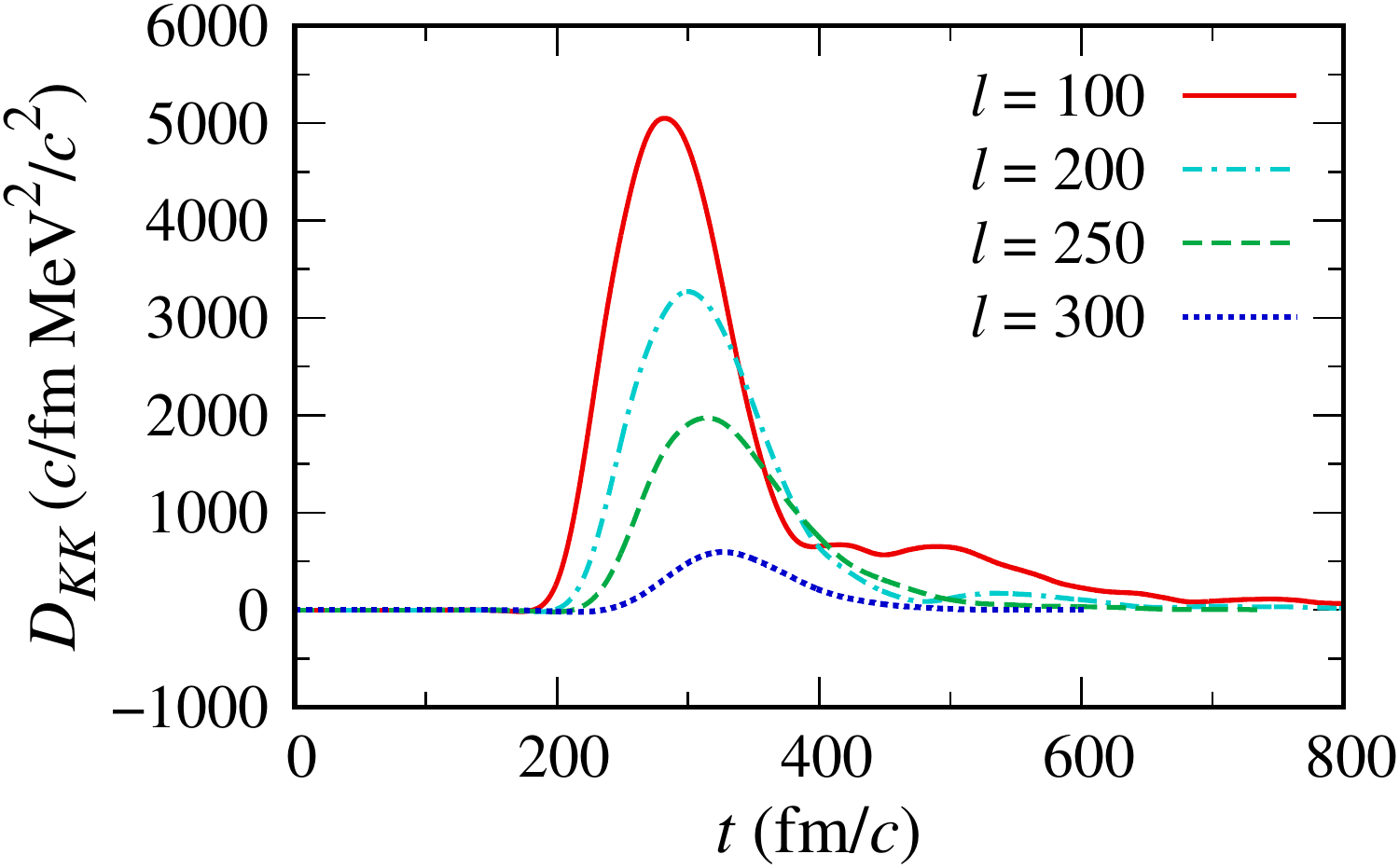}\vspace{-1mm}
\caption{
Radial-momentum diffusion coefficients in the $^{136}$Xe+$^{208}$Pb reaction
at \Ecm{526} with initial orbital angular momenta of $l$\,$=$\,100, 200, 250, and 300
(in units of $\hbar$) are shown as functions of time.
}\vspace{-3mm}
\label{fig:D_KK}
\end{figure}

\begin{figure} [t]
\includegraphics[width=8.6cm]{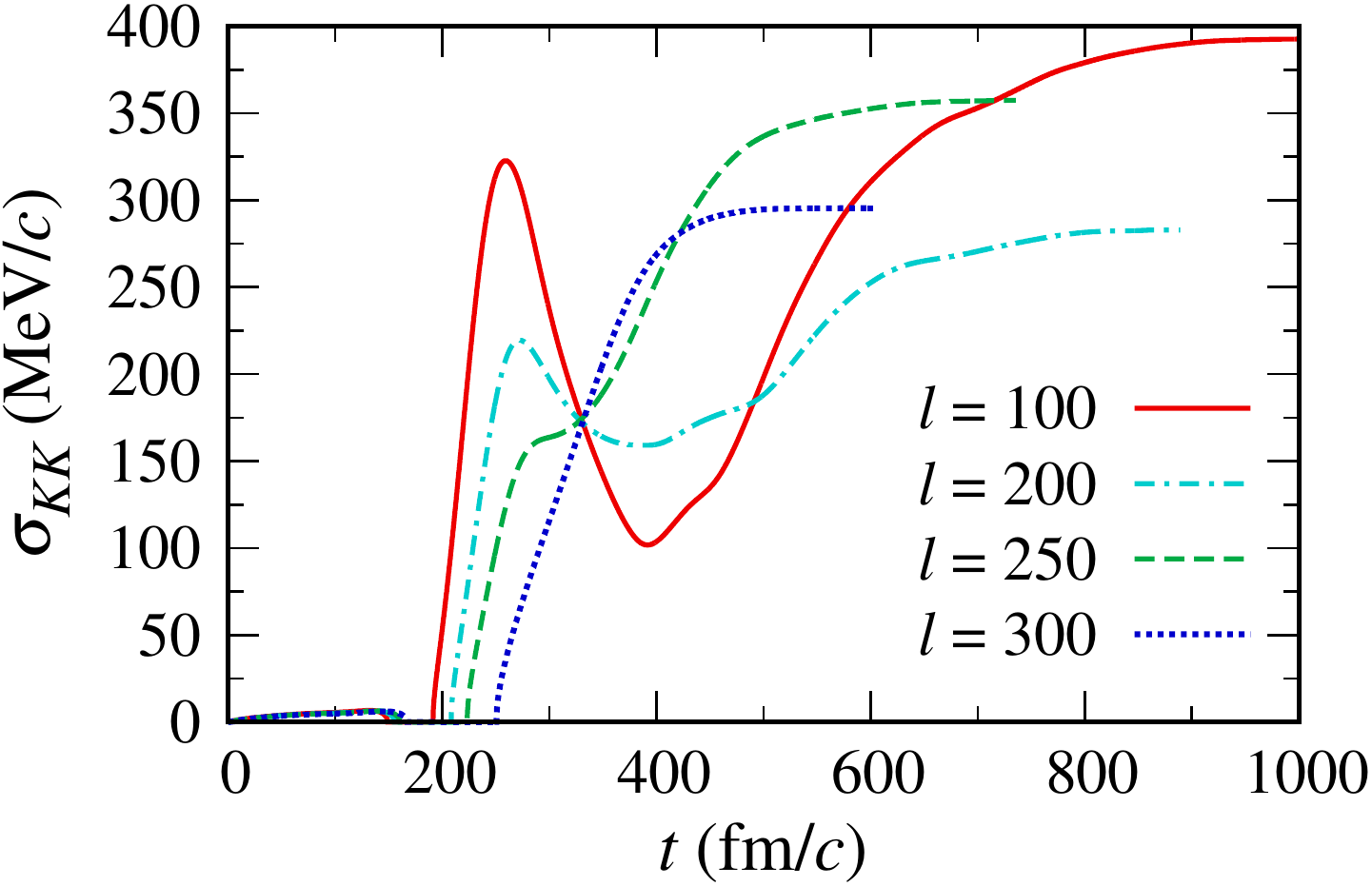}\vspace{-1mm}
\caption{
Variances of the radial momentum in the $^{136}$Xe+$^{208}$Pb reaction
at \Ecm{526} with initial orbital angular momenta of $l$\,$=$\,100, 200, 250, and 300
(in units of $\hbar$) are shown as functions of time.
}\vspace{-3mm}
\label{fig:sigma_KK}
\end{figure}

In Figs.~\ref{fig:gamma_R}--\ref{fig:sigma_KK}, we show examples of the
computational results for the collisions of $^{136}$Xe+$^{208}$Pb at \Ecm{526}
for four typical initial angular momenta, $l$ [$=$\,100 (solid line), 200 (dash-dotted line),
250 (dashed line), and 300 (dotted line) in units of $\hbar$], as functions of time.
Figure~\ref{fig:gamma_R} shows the reduced radial friction coefficients $\gamma_R(t,l)$
given by Eq.~(\ref{AEq:f_diss_2}), which were extracted from TDHF employing the
method explained in detail in Appendix~\ref{App:gamma_R}. We observe that the
radial friction coefficient develops when two nuclei collide at around $t$\,$=$\,200--400\,fm/$c$.
The magnitude of the friction coefficient increases with decreasing the initial orbital
angular momentum $l$, for which contact times are longer, indicating that larger
amount of the relative kinetic energy is converted into internal excitations at smaller
orbital angular momenta, as expected.
In Fig.~\ref{fig:D_KK}, we show the quantal momentum diffusion coefficient
$D_{KK}(t)$ given by Eq.~(\ref{Eq:D_alphabeta}), which is calculated microscopically
based on occupied single-particle orbitals within the TDHF approach. Again, the
magnitude of the diffusion coefficient increases with decreasing the initial orbital
angular momentum $l$. From the results, we find that the diffusion coefficient has
a relatively long tail as compared to the friction coefficient shown in Fig.~\ref{fig:gamma_R}.
It is related to the fact that the quantal diffusion coefficient is governed by nucleon
exchange which lasts even after the turning point through a neck structure of
the dinuclear system (cf.\ contact times shown in Table~\ref{Table:TDHF}).

Having the radial friction and momentum diffusion coefficients, $\gamma_R(t)$
and $D_{KK}(t)$, at hand, we solve the differential equation (\ref{Eq:PDE_sigma2_KK})
and the results are shown in Fig.~\ref{fig:sigma_KK}. From the figure, we see
that the variances of the radial momentum $\sigma_{KK}$ show somewhat
complicated behavior as a function of time. The variance grows in time and
saturates when two nuclei reseparate. We notice that the asymptotic value
of $\sigma_{KK}$ is largest for $l$\,$=$\,100$\hbar$ and decreases with
increasing $l$ values from 100$\hbar$ to 200$\hbar$, then increases for
$l$\,$=$\,250$\hbar$, and then decreases again for $l$\,$=$\,300$\hbar$.
One can also find this behavior in Table~\ref{Table:SMF}, in which the asymptotic
values of the radial momentum and TKE dispersions for a range of initial angular
momenta are presented. We consider that in the present analysis the radial
momentum dispersion is overpredicted for relatively large initial orbital angular
momentum region ($l$\,$=$\,(200--300)$\hbar$), which are probably due
to the approximate treatments of the radial friction coefficient. Since the primary
purpose of the present work is to put the first step toward the microscopic
description of the TKE distribution, developing a formalism based on the SMF
approach, we leave further improvements of the description as future works.

\begin{table}[t]
\caption{
A list of numerical results of the SMF calculations for the $^{136}$Xe+$^{208}$Pb reaction at \Ecm{526}
for a range of initial orbital angular momenta $l$. From left to right columns, it shows the asymptotic values of:
the radial momentum dispersion, $\sigma_{KK}$, in MeV/$c$, the modified radial momentum dispersion,
$\widetilde{\sigma}_{KK}=\sigma_{KK}/\sqrt{2\mu}$, in MeV$^{1/2}$, and the dispersion of total
kinetic energy (TKE), $\sigma_{\rm TKE}\approx2\tilde{\sigma}_{KK}\sqrt{E_{\rm kin}^\infty}$, in MeV.
}\vspace{-1mm}
\label{Table:SMF}
\begin{center}
\begin{tabular*}{\columnwidth}{@{\extracolsep{\fill}}cccc}
\hline\hline
$l$ ($\hbar$) & $\sigma_{KK}$ (MeV/$c$) & $\widetilde{\sigma}_{KK}$ (MeV$^{1/2}$) & $\sigma_{\rm TKE}$ (MeV) \\
\hline
0&    	553.7&	1.406&	52.84 \\
50&  	497.2&	1.263&	47.18 \\
100&	392.7&	0.999&	37.38 \\
110&	367.3&	0.934&	35.01 \\
120&	342.4&	0.871&	32.64 \\
130&	319.4&	0.813&	30.44 \\
140&	298.3&	0.759&	28.43 \\
150&	277.9&	0.706&	26.48 \\
160&	268.7&	0.681&	25.65 \\
170&	266.0&	0.674&	25.45 \\
180&	261.0&	0.661&	25.00 \\
190&	265.4&	0.672&	25.47 \\
200&	282.9&	0.718&	27.23 \\
210&	302.8&	0.769&	29.37 \\
220&	317.0&	0.805&	31.03 \\
230&	329.7&	0.838&	32.55 \\
240&	344.2&	0.876&	34.38 \\
250&	357.5&	0.910&	36.22 \\
260&	366.6&	0.933&	37.75 \\
270&	365.5&	0.930&	38.25 \\
280&	352.0&	0.896&	37.58 \\
290&	330.3&	0.840&	36.02 \\
300&	295.4&	0.752&	32.84 \\
310&	233.6&	0.595&	26.41 \\
320&	152.0&	0.387&	17.46 \\
330&	85.8&	0.218&	9.94 \\
340&	34.5&	0.088&	4.01 \\
350&	36.2&	0.092&	4.21 \\
\hline\hline
\end{tabular*}\vspace{-3mm}
\end{center}
\end{table}

\begin{figure} [t]
\includegraphics[width=8.6cm]{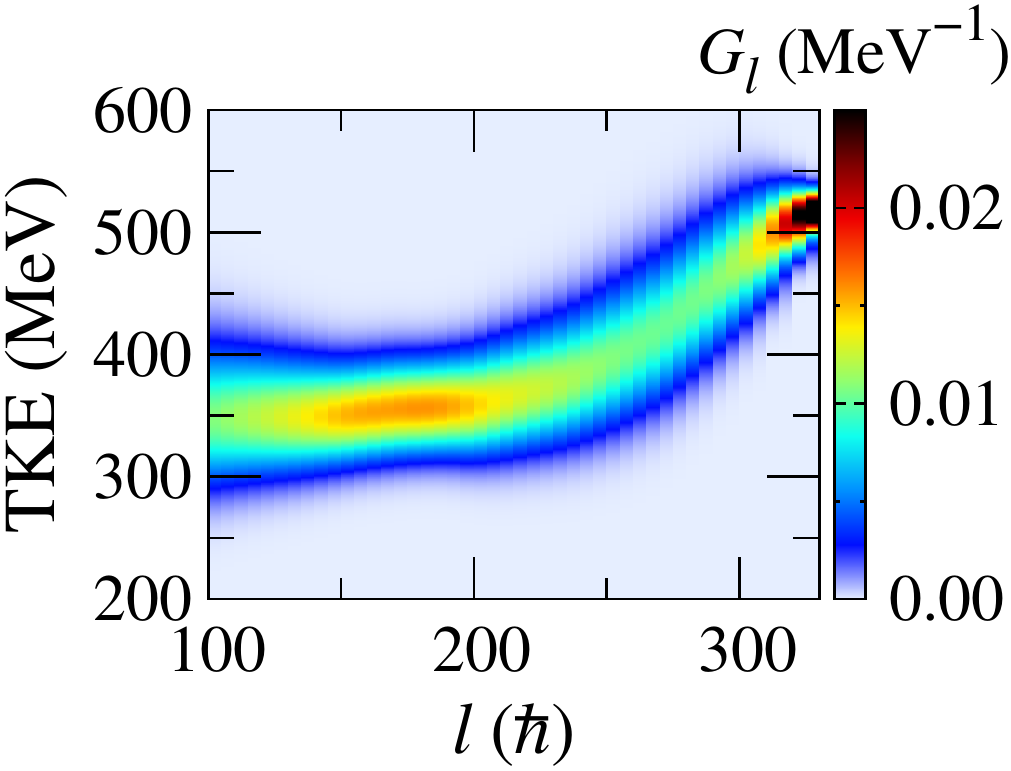}\vspace{-1mm}
\caption{
The total kinetic energy (TKE) distribution, $G_l$ defined by Eq.\,(\ref{Eq:TKE_dist}),
is shown in the $l$-TKE plane, where $l$ represents the initial orbital angular momentum
of the reaction, calculated for the $^{136}$Xe+$^{208}$Pb reaction at \Ecm{526}.
}\vspace{-3mm}
\label{fig:G_l}
\end{figure}

Employing the expression of Eq.~(\ref{Eq:TKE_dist}) we can obtain the TKE
distribution and the results are shown in Fig.~\ref{fig:G_l}. Figure~\ref{fig:G_l}
illustrates the TKE distribution for the range of initial angular momenta
$l$\,$=$\,(100--350)$\hbar$ in the $l$-TKE plane. We note that with
the TKE distribution, $G_l$, we can evaluate the mean value of TKE as
\begin{eqnarray}
\overline{{\rm TKE}(l)} &=& \int dE\, E\,G_l(E) \nonumber\\[1mm]
&\approx& E_{\rm kin}^\infty(l) + \widetilde{\sigma}_{KK}^2(l).
\end{eqnarray}
From Table~\ref{Table:SMF}, we see that the asymptotic value of
the largest dispersion occurs for $l=0$. In this case, we find
$\widetilde{\sigma}_{KK}^2=\sigma_{KK}^2/2\mu\approx1.98$\,MeV,
which is much smaller than $E_{\rm kin}^\infty(l)\approx353$\,MeV,
confirming the correspondence with the mean TKE from TDHF. We can
also calculate the variance of TKE for each value of angular momentum as
\begin{eqnarray}
\sigma_{\rm TKE}^2(l) &=& \int dE \bigl( E - E_{\rm kin}^\infty(l) \bigr)^2 G_l(E) \nonumber\\[1mm]
&\approx& 4\widetilde{\sigma}_{KK}^2(l)E_{\rm kin}^\infty(l).
\end{eqnarray}
Dispersion of TKE grows linearly with the square root of the mean value,
$\sigma_{\rm TKE}\approx2\widetilde{\sigma}_{KK}\sqrt{E_{\rm kin}^\infty}$.
For example for the initial angular momentum $l=0$, dispersion is as large as
$\sigma_{\rm TKE}\approx53$\,MeV. This indicates the total excitation
energy of the primary fragments have quite large dispersion values. Large
values of dispersions of the excitation energies may have an important effect
on deexcitation processes of the primary fragments.

Finally, to make a comparison with the experimental data \cite{Kozulin(2012)},
we evaluate the yield of the reaction outcomes as a function of total kinetic energy
loss (TKEL, i.e., $E_{\rm c.m.}-E_{\rm kin}^\infty$) by summing up contributions
from each initial orbital angular momentum,
\begin{equation}
Y(E_{\rm c.m.}-E_{\rm kin}^\infty) = Y_0\sum_{l=100}^{350}(2l+1)G_l(E_{\rm kin}^\infty).
\label{Eq:TKEL_dist}
\end{equation}
The normalization constant $Y_0$ is adjusted to the data at a suitable point.
The experimental setup in the work of Kozulin \textit{et al.}~\cite{Kozulin(2012)}
has an energy resolution of 25\,MeV. To compare with the data, this experimental
uncertainty should be accounted for by, e.g., a folding procedure of the calculated kinetic
energy distribution. The folding procedure will introduce approximately a uniform
shift in the kinetic energy distribution. Therefore, we consider that it does not
change the shape of the calculated curve and the folding effect is absorbed
by the normalization constant $Y_0$.

\begin{figure} [t]
\includegraphics[width=8.6cm]{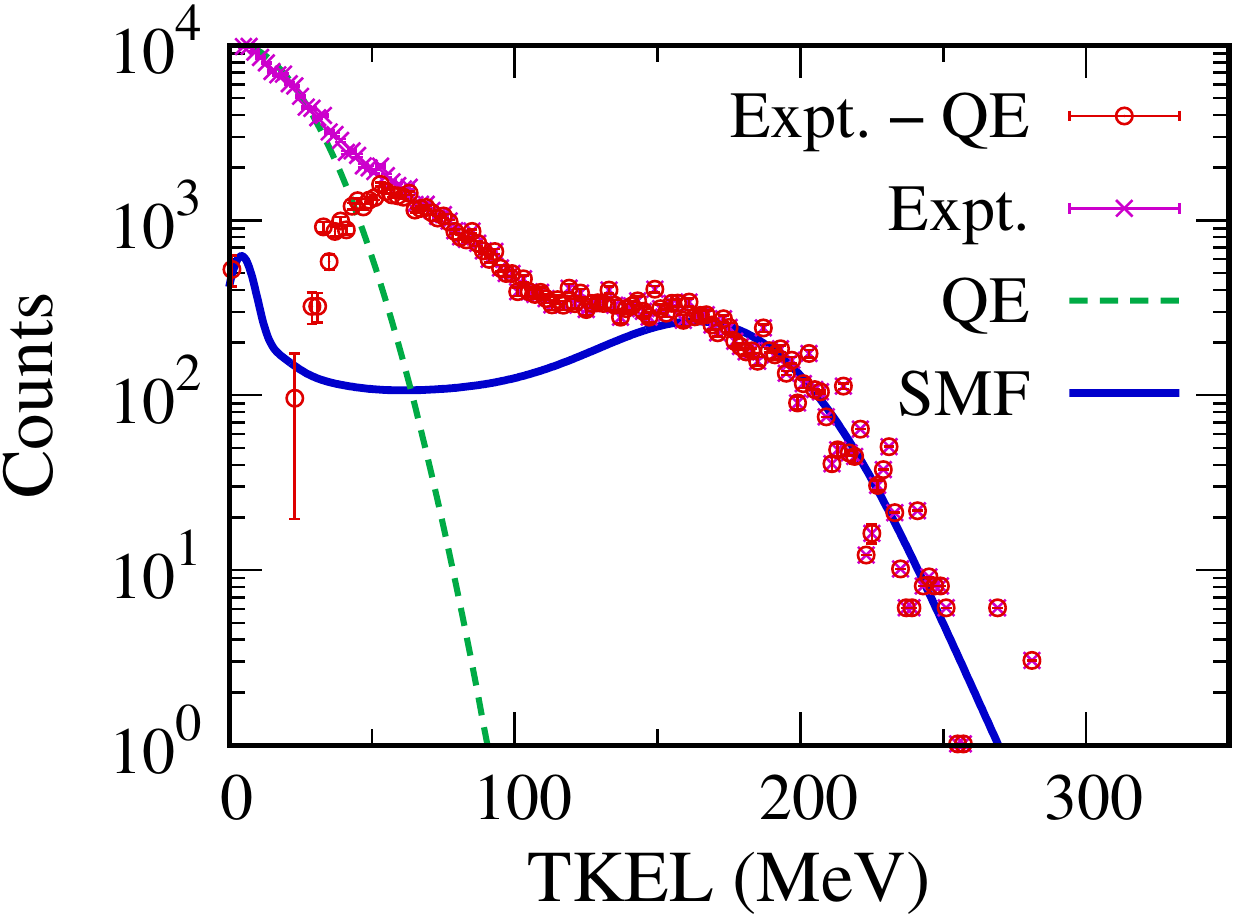}\vspace{-1mm}
\caption{
The integrated total kinetic energy loss (TKEL) distribution for the
$^{136}$Xe+$^{208}$Pb reaction at $E_{\rm c.m.}$\,=\,526\,MeV.
Magenta crosses (open circles) with error bars represent the experimental
data with (without) quasielastic contribution reported in Ref.~\cite{Kozulin(2012)}.
A Gaussian fit of the quasielastic contribution is shown by the green dashed
line. The results of SMF calculations are shown by the blue solid line, where
the normalization constant is set to $Y_0=3.5$.
}\vspace{-3mm}
\label{fig:TKE_Counts}
\end{figure}

Figure~\ref{fig:TKE_Counts} shows a comparison of the calculations with
experimental data for the collisions of $^{136}$Xe+$^{208}$Pb at \Ecm{526}.
The measured TKEL distribution for two-body events (without sequential fission
events) is shown by crosses with error bars. Open circles with error bars represent
the experimental data from which the quasielastic component (a Gaussian fit to the data,
shown by a dashed line) has been removed \cite{Kozulin(2012)}. The calculated
TKEL distribution according to Eq.~(\ref{Eq:TKEL_dist}) is shown by a solid line.
From the figure we find that the calculations provide good description for strongly-damped
events with large energy losses, TKEL\,$\gtrsim$\,150\,MeV. However, it underestimates
the count curve over the lower energy-loss segment. This behavior is a result of apparent
large dispersions of the TKE distribution over the range $l$\,$=$\,200$\hbar$--300$\hbar$,
which may be due to the over prediction of the radial momentum diffusion coefficients
and/or the approximate description of the radial friction for the large angular momentum
region. Although further improvements of the formalism are mandatory, we consider
that the quantal diffusion approach based on the SMF theory provides a promising
microscopic basis for quantifying kinetic energy dissipation and fluctuations in
low-energy heavy-ion reactions.

\section{Summary}\label{Sec:Conclusions}

The stochastic mean-field (SMF) approach goes beyond the standard
mean-field approximation, describing dynamical fluctuations of the
collective motion in heavy-ion collisions at low energies. In the time-dependent
Hartree-Fock (TDHF) approach, dynamical evolution of the colliding system
is described by a single Slater determinant which is determined by a given
set of initial conditions. In the SMF approach, on the other hand, an ensemble
of the mean-field (TDHF) events for stochastically-generated initial conditions
is considered. The initial conditions for each event are specified by the quantal
and thermal fluctuations, and each event evolves according to the self-consistent
mean-field Hamiltonian of that event. As a result, the SMF approach provides
not only the mean values, but the entire distribution functions of the observables.

For low-energy heavy-ion reactions in which the colliding system maintains
a dinuclear structure, instead of generating an ensemble of stochastic mean-field
events, the evolution of the system can be described in terms of a few representative
macroscopic variables, such as the relative linear momentum, the orbital angular momentum,
the mass and charge asymmetries of the colliding system. In such a case, we can deduce
effective equations of motion for the macroscopic variables by adiabatic or geometric
projection of the stochastically-generated mean-field events on a macroscopic subspace,
in a manner similar to the Mori formalism \cite{Mori(1965)}. For deep-inelastic or
quasifission processes in which dinuclear structure is maintained, the geometric
projection with the help of the window dynamics is more suitable to deduce the
effective equations for the macroscopic variables. Being consistent with the Mori formalism,
the effective equations for the macroscopic variables take the form of generalized
Langevin equations which offer quantal diffusion description of the dynamical evolution
of the colliding system. The Langevin equations are characterized by transport coefficients,
i.e., diffusion and drift coefficients. It is possible to deduce analytical expressions for transport
coefficients by carrying out suitable averages over the ensemble generated by the SMF approach.
Employing the closure relation in the diabatic approximations of the TDHF wave functions,
we can express the diffusion coefficients in terms of the occupied single-particle orbitals in
TDHF. Therefore, it provides a very practical and powerful framework for the microscopic
description of fluctuations of collective variables. This result is consistent with the quantal
fluctuation-dissipation theorem of the non-equilibrium statistical mechanics \cite{Gardiner(1991),
Weiss(1999)}. The theorem states that the diffusion coefficients which provide the
source of fluctuations can be expressed in terms of the mean-field properties.

In previous studies we employed the quantal diffusion approach to investigate
multinucleon transfer mechanism in low-energy heavy-ion collisions.  In the
present work, we have developed a formalism for describing the total kinetic
energy (TKE) distributions of binary reaction products. We have deduced an
effective transport equation for the relative linear momentum based on the
SMF approach by the projection technique with the help of window dynamics.
The radial and the tangential components of this equation provide quantal diffusion
description of the radial linear momentum and the angular momentum of the
relative motion, respectively.

As the first application of the present formalism,
we have analyzed the total kinetic energy loss (TKEL) distribution for the
$^{136}$Xe+$^{208}$Pb reaction at \Ecm{526}. For the analysis of the
TKEL distribution, in addition to the radial momentum diffusion coefficient,
it is necessary to determine the radial friction coefficient for different initial
angular momenta. The one-body dissipation mechanism is contained in the
mean-field description of the TDHF approach, but it is not trivial how to deduce
the expression for the radial friction force and the radial friction coefficient
from the TDHF description.  We inferred an approximate expression for the
radial friction force by using the analogy to the Langevin description of the random
walk problem. We find that the dispersion of the TKE distribution reach rather large
values which may have important effects in the deexcitation mechanism of the
primary fragments. We have calculated the TKEL distribution by summing over
the range of the initial angular momenta which is consistent with the experimental
angular coverage. We have found that the calculations provide a reasonable
description of the experimental data of the TKEL distribution for strongly-damped
events with large energy losses (TKEL\,$\gtrsim$\,150\,MeV), but underestimate
the data for lower values of TKEL (i.e., reactions at large initial orbital angular momenta).
The underestimation of the TKEL distribution for large values of the orbital
angular momentum may be partly due to the approximate description of
the radial friction coefficient and/or the neglected coupling between the
radial and tangential components of the linear momentum. This work
put an important step forward for the microscopic description of low-energy
heavy-ion reactions, including distributions of various observables, and
further improvements of the proposed formalism are in order.

\begin{acknowledgements}
The authors are grateful to Prof. E.M.~Kozulin for providing us the experimental data.
S.A. is very much thankful to F.~Ayik for continuous support and encouragement.
This work is supported in part by U.S. Department of Energy (DOE) Grant No.\ DE-SC0015513,
and in part by JSPS Grant-in-Aid for Early-Career Scientists Grant No.\ 19K14704.
This work used the computational resource of the HPCI system (Oakforest-PACS)
provided by Joint Center for Advanced High Performance Computing (JCAHPC)
through the HPCI System Project (Project ID: hp200022). This work also used
(in part) computational resources of the Cray XC40 System at Yukawa Institute
for Theoretical Physics (YITP), Kyoto Univesity.
\end{acknowledgements}

\appendix

\setcounter{figure}{0}
\renewcommand{\thefigure}{\thesection\arabic{figure}}

\section{Momentum diffusion coefficient}\label{App:A}

In this Appendix, we derive the quantal expression of the momentum
diffusion coefficient given by Eq.~(\ref{Eq:D_alphabeta}). Employing
a partial integration in the expression $Y_{ha}^\alpha(t)$ in
Eq.~(\ref{Eq:Y}), we have
\begin{eqnarray}
Y_{ha}^\alpha(t) &=& \frac{\hbar^2}{m}\int d\bs{r}_1 \Bigl[
g(x_1')[\hat{\bs{e}}_\alpha\bs{\cdot}\nabla_1\hat{\bs{e}}_R\bs{\cdot}\nabla_1\phi_h^*(\bs{r}_1,t)] \nonumber\\
&&+ \frac{1}{2}\hat{\bs{e}}_\alpha\bs{\cdot}\nabla_1g(x_1')\hat{\bs{e}}_R\bs{\cdot}\nabla_1\phi_h^*(\bs{r}_1,t) \nonumber\\
&&+ \frac{1}{2}\hat{\bs{e}}_R\bs{\cdot}\nabla_1g(x_1')\hat{\bs{e}}_\alpha\bs{\cdot}\nabla_1\phi_h^*(\bs{r}_1,t) \nonumber\\
&&+ \frac{1}{4}[\hat{\bs{e}}_\alpha\bs{\cdot}\nabla_1\hat{\bs{e}}_R\bs{\cdot}\nabla_1g(x_1')]\phi_h^*(\bs{r}_1,t)
\Bigr]\phi_a(\bs{r}_1,t), \nonumber\\[-1mm]
\end{eqnarray}
and its complex conjugation,
\begin{eqnarray}
Y_{ha}^{\alpha*}(t) &=& \frac{\hbar^2}{m}\int d\bs{r}_1 \Bigl[
g(x_1')[\hat{\bs{e}}_\alpha\bs{\cdot}\nabla_1\hat{\bs{e}}_R\bs{\cdot}\nabla_1\phi_h(\bs{r}_1,t)] \nonumber\\
&&+ \frac{1}{2}\hat{\bs{e}}_\alpha\bs{\cdot}\nabla_1g(x_1')\hat{\bs{e}}_R\bs{\cdot}\nabla_1\phi_h(\bs{r}_1,t) \nonumber\\
&&+ \frac{1}{2}\hat{\bs{e}}_R\bs{\cdot}\nabla_1g(x_1')\hat{\bs{e}}_\alpha\bs{\cdot}\nabla_1\phi_h(\bs{r}_1,t) \nonumber\\
&&+ \frac{1}{4}[\hat{\bs{e}}_\alpha\bs{\cdot}\nabla_1\hat{\bs{e}}_R\bs{\cdot}\nabla_1g(x_1')]\phi_h(\bs{r}_1,t)
\Bigr]\phi_a^*(\bs{r}_1,t). \nonumber\\[-1mm]
\end{eqnarray}
With the diabatic property of the TDHF wave functions, we can shift the wave
functions back and forth during short time intervals $\tau=t-t'$ to have an
approximate relation,
\begin{equation}
\phi_a(\bs{r},t') \approx \phi_a(\bs{r}-\bs{u}\tau,t),
\end{equation}
where $\bs{u}\tau$ denotes a small displacement during the short time interval
with flow velocity $\bs{u}$. Using the closure relation,
\begin{equation}
\sum_a\phi_a^*(\bs{r}_1,t)\phi_a(\bs{r}_2-\bs{u}\tau,t) = \delta(\bs{r}_1-\bs{r}_2+\bs{u}\tau),
\end{equation}
we obtain:
\begin{eqnarray}
\sum_{h\in T,a\in P}Y_{ha}^\alpha(t) Y_{ha}^{\beta*}(t')
&=& \sum_{h\in T}\iint d\bs{r}_1d\bs{r}_2\,\delta(\bs{r}_1-\bs{r}_2+\bs{u}_h\tau) \nonumber\\[-2mm]
&&\hspace{8mm}\times\;W_h^\alpha(\bs{r}_1,t)W_h^{\beta*}(\bs{r}_2,t).
\label{AEq:YY}
\end{eqnarray}
First, we consider the case for $\alpha$\,$=$\,$\beta$\,$=$\,$R$.
The radial part $W_h^R(\bs{r}_1,t)$ reads
\begin{eqnarray}
W_h^R(\bs{r}_1,t) &=& \frac{\hbar^2}{m}\Bigl[
g(x_1')[(\hat{\bs{e}}_R\bs{\cdot}\nabla_1)(\hat{\bs{e}}_R\bs{\cdot}\nabla_1)\phi_h(\bs{r}_1,t)] \nonumber\\
&&+ \frac{1}{2}\hat{\bs{e}}_R\bs{\cdot}\nabla_1g(x_1')\hat{\bs{e}}_R\bs{\cdot}\nabla_1\phi_h(\bs{r}_1,t) \nonumber\\
&&+ \frac{1}{2}\hat{\bs{e}}_R\bs{\cdot}\nabla_1g(x_1')\hat{e}_R\bs{\cdot}\nabla_1\phi_h(\bs{r}_1,t) \nonumber\\
&&+ \frac{1}{4}[(\hat{\bs{e}}_R\bs{\cdot}\nabla_1)(\hat{\bs{e}}_R\bs{\cdot}\nabla_1)g(x_1')]\phi_h(\bs{r}_1,t)
\Bigr]. \nonumber\\[-1mm]
\end{eqnarray}
Using the expression for $g(x')$ given by Eq.~(\ref{Eq:g}), we find
\begin{eqnarray}
W_h^K(\bs{r}_1,t)
&=& \frac{\hbar^2}{m}g(x_1')\Bigl[
[(\hat{\bs{e}}_R\bs{\cdot}\nabla_1)^2\phi_h(\bs{r}_1,t)] \nonumber\\
&&- \frac{x_1'}{\kappa^2}[\hat{\bs{e}}_R\bs{\cdot}\nabla_1\phi_h(\bs{r}_1,t)] \nonumber\\
&&+ \frac{1}{4\kappa^4}[x_1'^2-\kappa^2]\phi_h(\bs{r}_1,t),
\Bigr]
\end{eqnarray}
and its complex conjugation,
\begin{eqnarray}
W_h^{K*}(\bs{r}_2,t)
&=& \frac{\hbar^2}{m}g(x_2')\Bigl[
[(\hat{\bs{e}}_R\bs{\cdot}\nabla_2)^2\phi_h^*(\bs{r}_2,t)] \nonumber\\
&&- \frac{x_2'}{\kappa^2}[\hat{\bs{e}}_R\bs{\cdot}\nabla_2\phi_h^*(\bs{r}_2,t)] \nonumber\\
&&+ \frac{1}{4\kappa^4}[x_2'^2-\kappa^2]\phi_h^*(\bs{r}_2,t).
\Bigr]
\end{eqnarray}
Let us introduce the following coordinate transformations,
\begin{equation}
\bs{R} = (\bs{r}_1+\bs{r}_2)/2, \hspace{5mm}
\bs{r} = \bs{r}_1-\bs{r}_2,
\end{equation}
and its inverse,
\begin{equation}
\bs{r}_1 = \bs{R}+\bs{r}/2, \hspace{5mm}
\bs{r} = \bs{R} - \bs{r}/2.
\end{equation}
Because of the delta function, we can immediately carry out
the integration over $\bs{r}$ in Eq.~(\ref{AEq:YY}) and make
the substitution for $\bs{r}=-\bs{u}_h\tau$, and introduce
diabatic shifts in the wave functions,
\begin{eqnarray}
\phi_h(\bs{r}_1,t) &=& \phi_h(\bs{R}+\bs{r}/2,t)
= \phi_h(\bs{R}-\bs{u}_h\tau/2,t) \nonumber\\[1mm]
&\approx& \phi_h(\bs{R},\bar{t}), \\[2mm]
\phi_h(\bs{r}_2,t) &=& \phi_h(\bs{R}-\bs{r}/2,t)
= \phi_h(\bs{R}+\bs{u}_h\tau/2,t) \nonumber\\[1mm]
&\approx& \phi_h(\bs{R},\bar{t}),
\end{eqnarray}
with $\bar{t}\equiv(t+t')/2$. We can express product of the Gaussian factors as
$g(x_1')g(x_2')=\tilde{g}(X')\tilde{G}(x')$, where
\begin{eqnarray}
\tilde{g}(X') &=& \frac{1}{\sqrt{\pi}\kappa}\exp\biggl[-\biggl(\frac{X'}{\kappa}\biggr)^2\biggr], \\
\tilde{G}(X') &=& \frac{1}{\sqrt{4\pi}\kappa}\exp\biggl[-\biggl(\frac{X'}{2\kappa}\biggr)^2\biggr],
\end{eqnarray}
with $x'=\hat{\bs{e}}_R\bs{\cdot}\bs{r}=-\hat{\bs{e}}_R\bs{\cdot}\hat{\bs{u}}_h\tau=u_R^h\tau$
and $X'=\hat{\bs{e}}_R\bs{\cdot}\bs{R}$, where $u_R^h(\bs{R},\bar{t})$ denotes
the component of the flow velocity of the hole states perpendicular to the window,
which may, in general, depend on the mean position $\bs{R}=(\bs{r}_1+\bs{r}_2)/2$
and the mean time $\bar{t}=(t+t')/2$. In the product $W_h^\alpha(\bs{r}_1,t)W_h^{\beta*}(\bs{r}_2,t')$,
there are linear, second, third, and fourth order terms in $x_1$ and $x_2$ in the integrand
of Eq.~(\ref{AEq:YY}). The integrand contains a product of two sharp Gaussians, $\tilde{g}(X')$
and $\tilde{G}(x')$, which provides the memory kernel in the integrand. Taking the
averages over the memory kernel and over the sharp Gaussian $\tilde{g}(X')$,
all terms in the integrand of Eq.~(\ref{AEq:YY}) which are proportional to the
powers of $x_1$ and $x_2$ vanish. We obtain the similar results for other
components of Eq.~(\ref{AEq:YY}) with $\alpha,\beta=R,\theta$, and we find
\begin{eqnarray}
\sum_{h\in{\rm T}, a\in{\rm P}}Y_{ha}^\alpha(t)Y_{ha}^{\beta*}(t')
&=& \Bigl(\frac{\hbar^2}{m}\Bigr)^2 \sum_{h\in T}\int d\bs{R}\;\tilde{g}(X')
\frac{G_{\rm T}^h(\tau)}{|u_R^h(\bs{R},\bar{t})|} \nonumber\\[0.5mm]
&&\times\;\bigl[(\hat{\bs{e}}_\alpha\bs{\cdot}\nabla)(\hat{\bs{e}}_R\bs{\cdot}\nabla)\phi_h(\bs{R},\bar{t})\bigr] \nonumber\\[0.5mm]
&&\times\;\bigl[(\hat{\bs{e}}_\beta\bs{\cdot}\nabla)(\hat{\bs{e}}_R\bs{\cdot}\nabla)\phi_h(\bs{R},\bar{t})\bigr]^*, \nonumber\\
\label{AEq:YY_2}
\end{eqnarray}
where the memory kernel is defined as
\begin{equation}
G_{\rm T}^h(\tau) = \frac{1}{\sqrt{4\pi}}\frac{1}{\tau_{\rm T}^h}\exp\biggl[ -\biggl(\frac{\tau}{2\tau_{\rm T}^h}\biggr)^2 \biggr],
\label{AEq:G_T}
\end{equation}
with the memory time, $\tau_{\rm T}^h=\kappa/|u_R^h|$. We can write
the wave functions as $\phi_h(\bs{r},t)=|\phi_h(\bs{r},t)|\exp(iQ_h)$
\cite{Gottfried(1966)}. Since the phase factor behaves like the velocity
potential, neglecting derivative of the amplitude of the wave function,
we have the approximate result:
\begin{eqnarray}
(\hat{\bs{e}}_R\bs{\cdot}\nabla)\phi_h
&\approx& i\phi_h\,\hat{\bs{e}}_R\bs{\cdot}\nabla Q_h \nonumber\\[0.5mm]
&=& i\phi_h(\bs{R},\bar{t})\frac{m}{\hbar}u_R^h(\bs{R},\bar{t}).
\label{AEq:dphi}
\end{eqnarray}
In a similar manner, we can express the second derivative of the wave function as,
\begin{eqnarray}
(\hat{\bs{e}}_\theta\bs{\cdot}\nabla)(\hat{\bs{e}}_R\bs{\cdot}\nabla)\phi_h
&\approx& i[(\hat{\bs{e}}_\theta\bs{\cdot}\nabla)\phi_h(\bs{R},\bar{t})]u_R^h(\bs{R},\bar{t}) \nonumber\\[0.5mm]
&\approx& -\phi_h(\bs{R},\bar{t})\frac{m^2}{\hbar^2}u_\theta^h(\bs{R},\bar{t})u_R^h(\bs{R},\bar{t}). \nonumber\\
\label{AEq:ddphi}
\end{eqnarray}
We can write the radial ($\alpha$\,$=$\,$R$) and the tangential
($\alpha$\,$=$\,$\theta$) flow velocities in the flowing form,
\begin{eqnarray}
u_\alpha^h(\bs{R},\bar{t})
&=& \frac{\hbar}{m}\frac{{\rm Im}\bigl[ \phi_h^*(\bs{R},\bar{t})\hat{\bs{e}}_\alpha\bs{\cdot}\nabla\phi_h(\bs{R},\bar{t}) \bigr]}{|\phi_h(\bs{R},\bar{t})|^2}. \nonumber\\
\end{eqnarray}
Incorporating this expression, Eq.~(\ref{AEq:YY_2}) becomes,
\begin{eqnarray}
\sum_{h\in{\rm T},a\in{\rm P}}Y_{ha}^\alpha Y_{ha}^{\beta*}
= \sum_{h\in T}\int d\bs{R}\;\tilde{g}(X')G_{\rm T}(\tau)J_{\alpha\beta}^{\rm T}(\bs{R},\bar{t}). \nonumber\\[-2mm]
\end{eqnarray}
Here, $J_{\alpha\beta}^{\rm T}(\bs{R},\bar{t})$ is given in Eq.~(\ref{Eq:J_T}),
and $G_{\rm T}(\tau)$ denotes the average memory kernel given by Eq.~(\ref{AEq:G_T}),
which is evaluated with the average value of the flow velocity of the hole states originating
from the target. The second term on the right side of Eq.~(\ref{Eq:bar_dfdf}) is evaluated
in a similar manner, and we obtain the expression given by Eq.~(\ref{Eq:D_alphabeta})
for the momentum diffusion coefficients.

\section{Radial friction coefficient}\label{App:gamma_R}

In this Appendix, we discuss an analysis of the radial friction coefficient
based on the mean-field solution of TDHF. The TDHF description contains
the one-body dissipation of relative kinetic energy and the transfer of the
relative angular momentum into the intrinsic degrees of freedom. The
dominant mechanism for the one-body dissipation is nucleon exchange
between projectile-like and target-like fragments. However, a microscopic
derivation of the so-called window formula for the reduced friction coefficients
from the TDHF approach is not trivial. Here, we consider the analogy with
the random walk problem to deduce the reduced friction coefficients from
the mean description of TDHF. By taking the ensemble average in Eq.~(\ref{Eq:ROC_Prel}),
the mean evolution of the rate of change of the relative momentum
is given by
\begin{equation}
\frac{\partial}{\partial t}\bs{P} = \int d\bs{r}\,g(x')\,\dot{x}'m\bs{j}(\bs{r},t)
+ \mbox{[Potential terms]} + \bs{f}(t),
\end{equation}
where $\bs{j}(\bs{r},t)=\frac{\hbar}{m}\sum_h{\rm Im}[\phi_h^*(\bs{r},t)\nabla\phi_h(\bs{r},t)]$.
This equation is equivalent to the TDHF description of the relative momentum.
The first and the second terms on the right hand side are the conservative forces on the
relative motion due the motion of the window and the potential terms. In the last term
$\bs{f}(t)$ represents the dynamical force due to nucleon exchange between
the projectile- and target-like fragments with the radial and the tangential components,
\begin{eqnarray}
f_\alpha(t) = \int d\bs{r}\,g(x') \sum_h \hat{\bs{e}}_R\bs{\cdot} \Bigl( \bs{A}_{hh}^\alpha - \bs{B}_{hh}^\alpha \Bigr)
\end{eqnarray}
Using the approximate result of Eq.~(\ref{AEq:dphi}) in Appendix~\ref{App:A},
we can express the component of the dynamical force due to nucleon exchange as
\begin{eqnarray}
f_\alpha(t) &=& -\frac{\hbar}{m}\int d\bs{r}\,g(x')\,\sum_h mu_\alpha^h(\bs{r},t) \nonumber\\
&&\hspace{10mm}\times\;{\rm Im}\bigl[ \phi_h^*(\bs{r},t)\hat{\bs{e}}_R\bs{\cdot}\nabla\phi_h(\bs{r},t) \bigr],
\end{eqnarray}
where the summation runs over the hole states originating from both projectile
and target nuclei. The dynamical force involves both the conservative and dissipative
forces. In order to infer the dissipative part of the dynamical force, we use the analogy
to the Langevin description of the random walk problem. As seen in Eq.~(\ref{Eq:D_alphabeta}),
the direct terms of the momentum diffusion coefficients are determined by
the sum of nucleon fluxes that carry the product of momentum components
from projectile to target and vice versa. In analogy to the description
of the random walk, the components of the dissipative force are determined by
the net momentum flux across the window as follows:
\begin{eqnarray}
f_\alpha^{\rm diss}(t)
&=& -\frac{\hbar}{m}\int d\bs{r}\,g(x')\sum_{h\in{\rm P}}mu_\alpha^h(\bs{r},t) \nonumber\\[-1mm]
&&\hspace{10mm}\times\Bigl|{\rm Im}\bigl[\phi_h^*(\bs{r},t)\hat{\bs{e}}_R\bs{\cdot}\nabla\phi_h(\bs{r},t)\bigr]\Bigr| \nonumber\\
&& +\frac{\hbar}{m}\int d\bs{r}\,g(x')\sum_{h\in{\rm T}}mu_\alpha^h(\bs{r},t) \nonumber\\[-1mm]
&&\hspace{10mm}\times\Bigl|{\rm Im}\bigl[\phi_h^*(\bs{r},t)\hat{\bs{e}}_R\bs{\cdot}\nabla\phi_h(\bs{r},t)\bigr]\Bigr|.
\label{AEq:f_diss}
\end{eqnarray}
Here, we express the net momentum flux as the difference of the momentum
flux carried by the hole orbitals originating from the projectile and the momentum
flux carried by the hole orbitals originating from the target. This approximate
description may over estimate the net momentum flux, in particular in collisions
at large impact parameters, and may require improvements. The quantity
$\bigl|{\rm Im}[\phi_h^*(\bs{r},t)\hat{\bs{e}}_R\bs{\cdot}\nabla\phi_h(\bs{r},t)]\bigr|$
represents the magnitude of the nucleon flux from one fragment to the other.

\begin{figure} [t]
\includegraphics[width=8cm]{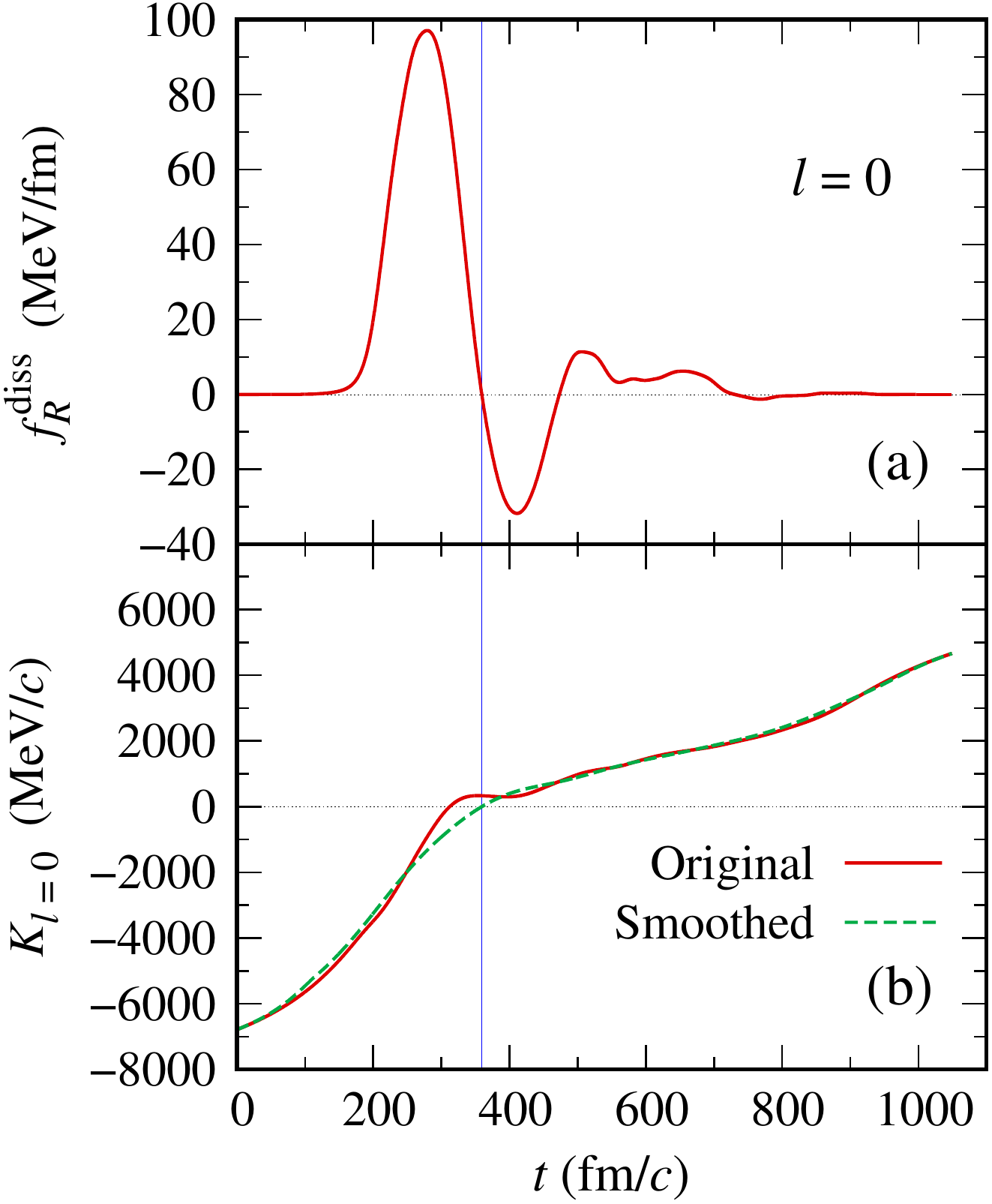}\vspace{-1mm}
\caption{
The radial friction force $f_{l=0}^{\rm diss}$ (a) and the radial momentum $K_{l=0}$ (b)
are shown as a function of time for the central collision ($l=0$). In panel (b),
the original value of the radial momentum obtained from TDHF is shown by the
solid line, while a smoothed curve is represented by the dashed line. The vertical line
indicates the time at which $f_{l=0}^{\rm diss}$ and the smoothed $K_{l=0}$ vanish.
}\vspace{-3mm}
\label{fig:B1}
\end{figure}

\begin{figure} [t]
\includegraphics[width=8cm]{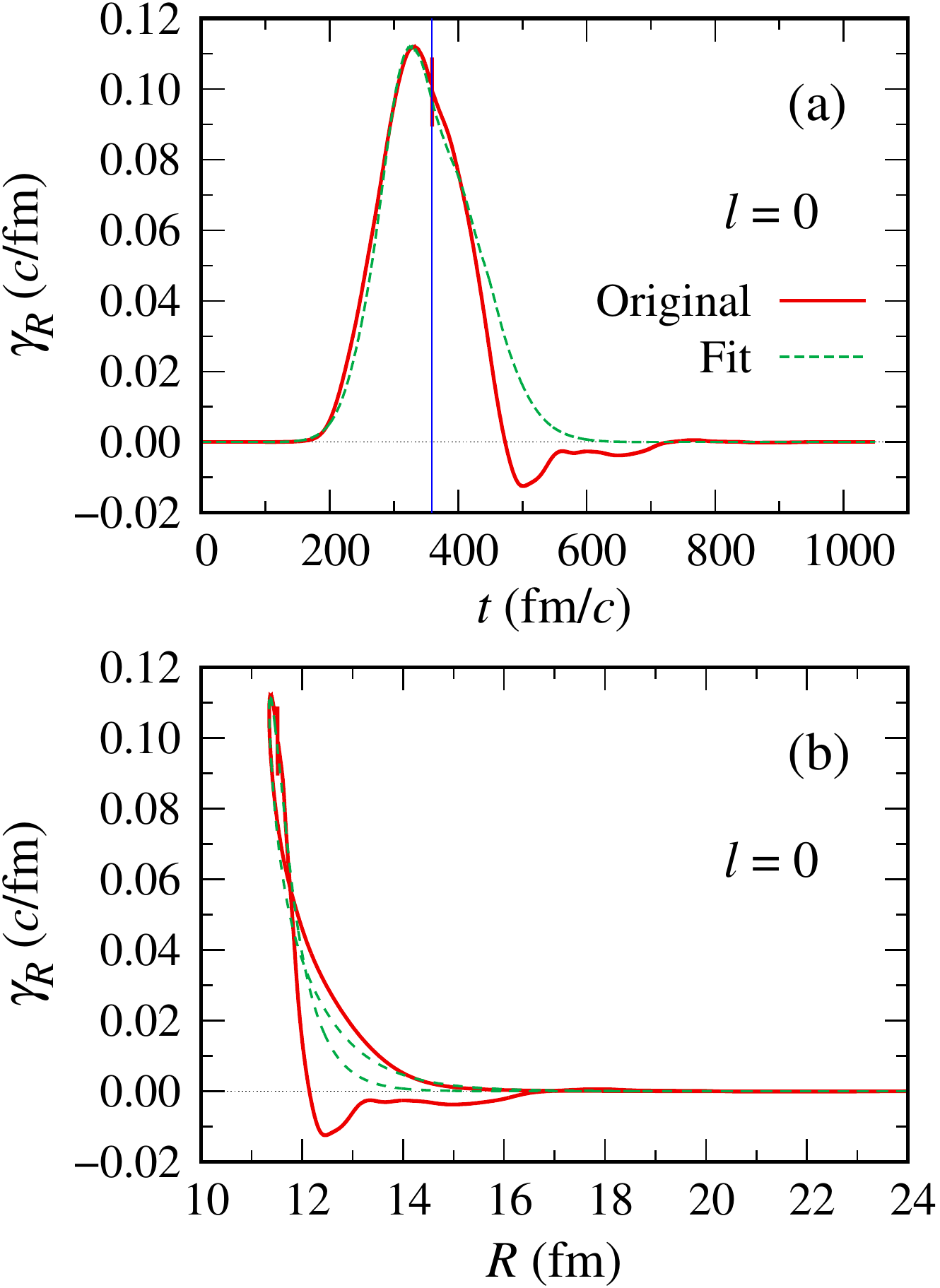}\vspace{-1mm}
\caption{
The extracted reduced radial friction coefficient, $\gamma_R(t,l=0)=-f_{l=0}^{\rm diss}(t)/\bar{K}_{l=0}(t)$,
is shown by the solid line for the central collision ($l=0$), where $\bar{K}_{l=0}(t)$ is the smoothed
radial momentum shown in Fig.~\ref{fig:B1}(b) by the dashed line. In panel (a), it is shown as a
function of time, while the same quantity is shown as a function of the relative distance in panel (b).
The dashed line represents the parametrized function given by Eqs.~(\ref{Eq:gamma_inc}) and (\ref{Eq:gamma_out}).
}\vspace{-3mm}
\label{fig:B2}
\end{figure}

In this work, we consider the radial friction force and the reduced radial friction
coefficient. In order to derive the expressions for the radial friction coefficient
$\gamma_R(t,l)$ for collisions with a range of initial angular momenta $l$,
we assume the phenomenological expression for the radial dissipative force,
\begin{equation}
f_R^{\rm diss}(t,l) = -\gamma_R(t,l)K_l(t),
\label{AEq:f_F_phenom}
\end{equation}
where $K_l=\hat{\bs{e}}_R\bs{\cdot}\bs{P}_l$ is the radial component of the
relative linear momentum. From this relation, in principle, it should be possible to
deduce the radial friction coefficient for each value of the initial angular momentum $l$.
In Fig.~\ref{fig:B1}, we show the radial friction force in (a) and the radial momentum
in (b) for the central collisions ($l=0$) as a function of time. The radial momentum vanishes
at the turning point which occurs at $t=313$\,fm/$c$ (see Fig.~\ref{fig:B1}(b), solid line).
We expect that the radial friction force also vanishes at the turning point. However, we notice
that the friction force vanishes at a slightly later time, $t=360$\,fm/$c$. The time shift may
originate from the approximate expression of the friction force, Eq.~(\ref{AEq:f_diss}),
which overestimates the net momentum flux across the window, and the shift becomes
larger for increasing the orbital angular momentum.

In order to extract the radial friction coefficients for all values of the initial angular
momentum $l$, we employ an approximate method as described below. In Fig.~\ref{fig:B1}(b),
we introduce a smoothing of the radial momentum so that the friction force and the
radial momentum vanish at the same instant. The smoothed (averaged over a time
interval of 220\,fm/$c$) radial momentum, say $\bar{K}_{l=0}(t)$, is shown by
the dashed line in Fig.~\ref{fig:B1}(b), and the vertical line indicates the instant
at which both the friction force and the radial momentum become vanishingly small.
Then, we define the reduced friction coefficient for $l=0$ as
\begin{equation}
\gamma_R(t,l=0) = -\frac{f_R^{\rm diss}(t,l=0)}{\bar{K}_{l=0}(t)}.
\label{AEq:gamma_R_phenom}
\end{equation}
Note that the phenomenological relation (\ref{AEq:f_F_phenom}) has been used.

\begin{figure} [t]
\includegraphics[width=8cm]{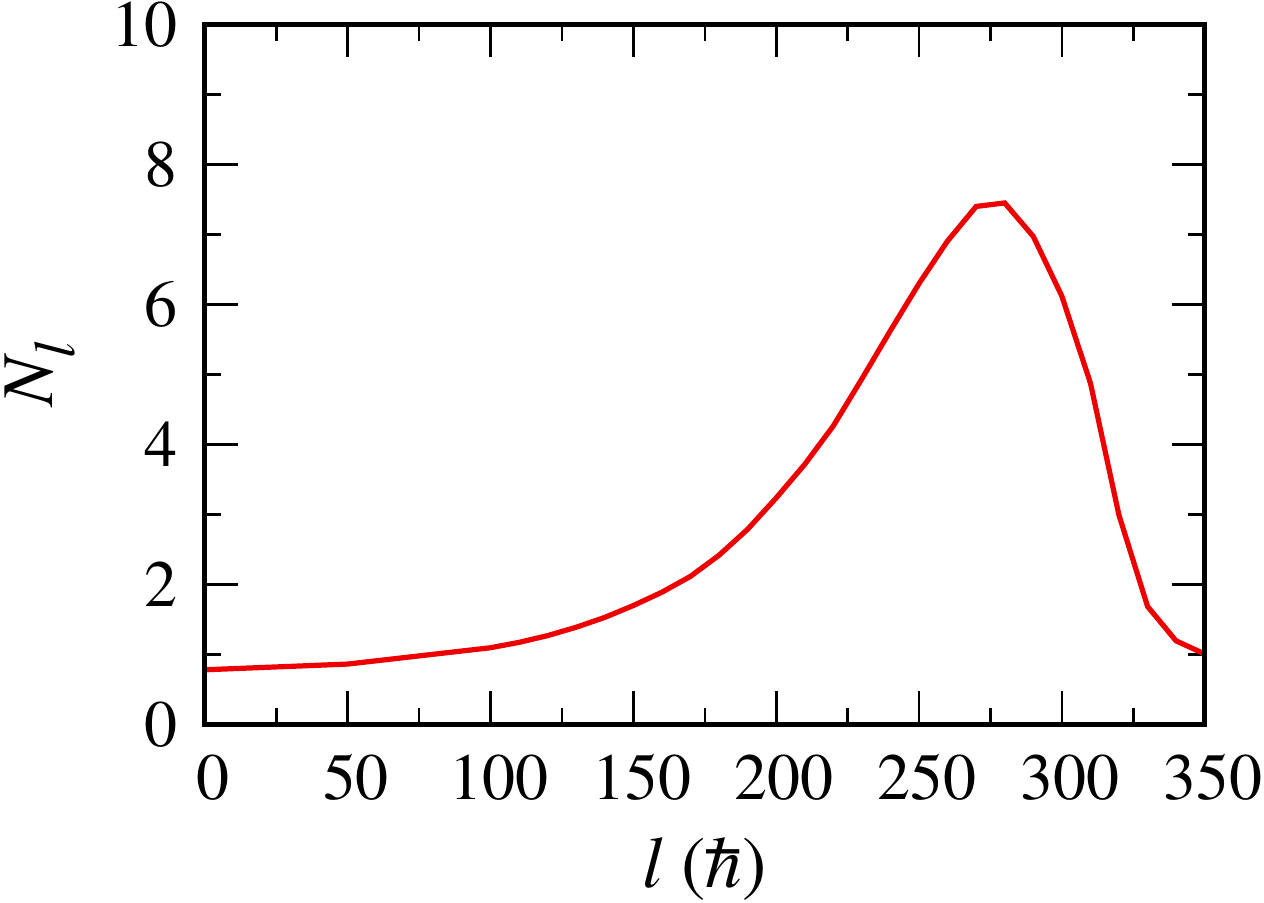}\vspace{-1mm}
\caption{
The normalization coefficient $N_l$ in Eq.~(\ref{AEq:f_diss_2}) is shown
as a function of the initial orbital angular momentum $l$.
}\vspace{-3mm}
\label{fig:B3}
\end{figure}

In Fig.~\ref{fig:B2}(a), we show the extracted friction coefficient according to
Eq.~(\ref{AEq:gamma_R_phenom}) in the central collision ($l=0$) as a function
of time (solid line). We find that dissipation occurs mainly during the incoming phase
until the turning point at around $t=360$\,fm/$c$ (indicated by a vertical line),
and only a small fraction of dissipation takes place during the outgoing phase after
the turning point until the separation of the fragments. We should thus ignore
an unphysical negative tail after $t=460$\,fm/$c$. Figure~\ref{fig:B2}(b) shows
the friction coefficient $\gamma_R$ in the central collision ($l=0$) as a function
of the relative distance $R(t)$. In this work, we parametrize the friction coefficient
during the incoming phase by an exponential function as
\begin{equation}
\gamma_R^{\rm inc}[R(t)] = c_1\exp\Bigl[ -c_2\bigl(R(t)-c_3\bigr) \Bigr],
\label{Eq:gamma_inc}
\end{equation}
with $c_1=9.97$\,$c$/fm, $c_2=1.04$\,fm$^{-1}$, and $c_3=6.86$\,fm.
The friction force reaches the maximum value at the minimum distance $R_{\rm min}$.
In the outgoing phase, from the minimum distance $R_{\rm min}$ to reseparation,
we adopt another form with an exponential damping factor as
\begin{equation}
\gamma_R^{\rm out}[R(t)] = \gamma_R^{\rm inc}[R(t)]\,c_4\exp\Bigl[ -c_5\bigl(R(t)-R_{\rm min}\bigr) \Bigr],
\label{Eq:gamma_out}
\end{equation}
where $c_4=1.96$ and $c_5=0.95$. Note that $c_4>1$ has been used,
because the minimum distance is reached at $t$\,$=$\,313\,fm/$c$,
while the obtained friction coefficient has a peak at slightly later time.
We joint the two expressions for the incoming and outgoing phases
smoothly around the turning point.
Assuming
that the friction coefficients scale with the relative distance for all initial
angular momentum in a similar manner as for the central collision, we
express the reduced radial friction coefficient for non-zero $l$ values as,
\begin{equation}
\gamma_R(t,l) = N_l\,\gamma_R[R_l(t)].
\label{AEq:f_diss_2}
\end{equation}
Here, $\gamma_R(R)$ is the friction coefficients given by Eqs.~(\ref{Eq:gamma_inc})
and (\ref{Eq:gamma_out}) extracted from the $l$\,$=$\,0 case as a function
of the relative distance. The normalization factor $N_l$ is determined by matching
the dissipated energy with the mean TKEL calculated by TDHF for each initial
angular momentum, i.e.,
\begin{equation}
E_l^{\rm diss} = \int dt\,\gamma_R(t,l)\frac{K_l^2(t)}{\mu(t)} = E_{\rm kin}^\infty(l).
\label{AEq:E_diss}
\end{equation}
In Fig.~\ref{fig:B3}, we show the magnitude of the normalization constant
$C_l$ as a function of the initial orbital angular momentum $l$. Figure~\ref{fig:gamma_R}
in the main text presents the reduced radial friction coefficients determined in the
manner outlined above as a function of time for typical initial orbital angular momenta.

\begin{figure} [t]
\includegraphics[width=8.6cm]{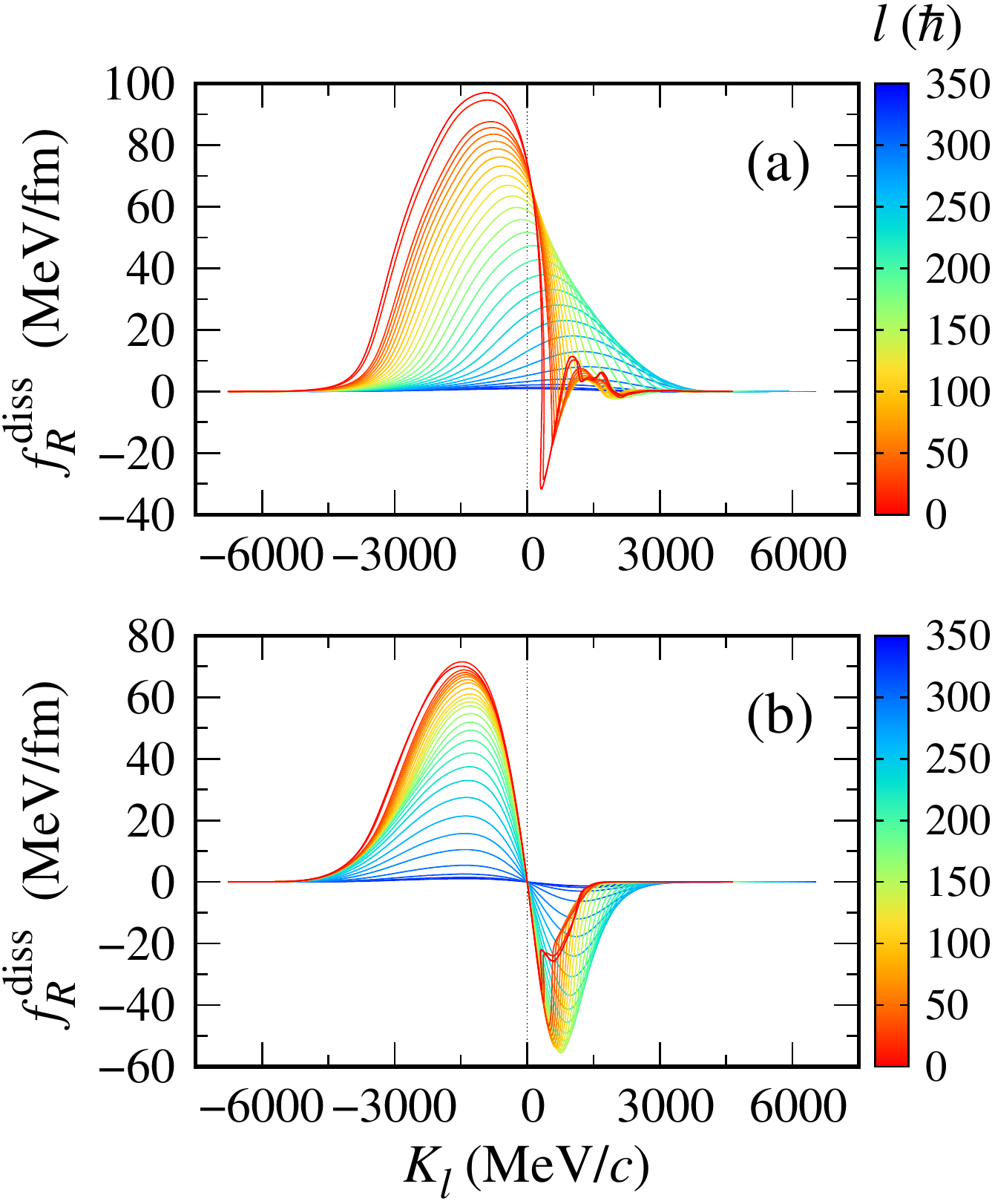}\vspace{-1mm}
\caption{
The dissipative force $F_l^{\rm diss}$ is shown as a function of the radial linear
momentum $K_l$ for the $^{136}$Xe+$^{208}$Pb reaction at \Ecm{526} with
a range of $l$ values. Colors represent the value of the initial orbital angular momenta.
(a) The dissipative force obtained with Eq.~(\ref{AEq:f_diss}) based on the single-particle
orbitals in TDHF. (b) The reconstructed dissipative force according to the phenomenological
expression, $F_l^{\rm diss}=-\gamma_RK_l$, where $\gamma_R$ here is expressed as
the fitted function given by Eqs.~(\ref{Eq:gamma_inc}) and (\ref{Eq:gamma_out}).
}\vspace{-3mm}
\label{fig:B4}
\end{figure}

In Fig.~\ref{fig:B4}, we compare the original radial friction force as given by
Eq.~(\ref{AEq:f_diss}) in panel (a) and the reconstructed radial friction
force using the approximate treatment of Eq.~(\ref{AEq:f_diss_2}) in panel
(b) as functions of the radial momentum for a range of initial angular momenta $l$.
It is visible that the original radial friction forces shown in (a) does not vanish at
the turning point at which the radial momentum changes its sign. On the other
hand, the reconstructed friction forces as functions of the radial momentum shown
in (b) give rise to the expected behavior and provide a support for the reduced
friction coefficients that we obtained using the approximate procedure.

\setcounter{figure}{0}
\section{Comparison with DD-TDHF}\label{App:DD-TDHF}

In this Appendix, we provide a supplemental analysis of the radial friction coefficient
for head-on collision ($l=0$) based on an alternative approach, called dissipative-dynamics
TDHF (DD-TDHF). The idea of DD-TDHF was first proposed in 1980 by Koonin \cite{DDTDHF0},
which was later tested for realistic applications by Lacroix in 2002 \cite{DDTDHF0.5}
and further applied by Washiyama \textit{et al.} \cite{DDTDHF1,DDTDHF2,DDTDHF3}.
(See Ref.~\cite{DDTDHF-review} for a short review.) Here let us succinctly
recall the basic idea of DD-TDHF.

In DD-TDHF, we consider a mapping of TDHF dynamics onto a set of classical
equations of motion:
\begin{eqnarray}
\frac{dR}{dt} &=& \frac{P}{\mu},\\
\frac{dP}{dt} &=& -\frac{dV_{\rm DD}}{dR} -\frac{d}{dR}\biggl(\frac{P^2}{2\mu}\biggr) -\gamma_R P,
\end{eqnarray}
where $V_{\rm DD}(R)$ and $\gamma_R(R)$ denote the nucleus-nucleus
potential and the reduced radial friction coefficient, respectively, as a function
of the relative distance $R$. A standard TDHF simulation provides time evolution
of the relative distance $R(t)$, the relative linear momentum $P(t)$, and the
reduced mass $\mu(t)$. Assuming that a slight change of collision energy
does not affect the two unknown quantities, i.e., $V_{\rm DD}(R)$ and $\gamma_R(R)$,
one can solve the above equations for them. Namely, one finds:
\begin{eqnarray}
\frac{dV_{\rm DD}(R)}{dR}
&=&
\frac{\dot{R}_{\rm II}\dot{P}_{\rm I}-\dot{R}_{\rm I}\dot{P}_{\rm II}}{\dot{R}_{\rm I}-\dot{R}_{\rm II}}
-\frac{1}{2}\frac{d\mu}{dR}\dot{R}_{\rm I}\dot{R}_{\rm II},
\label{Eq:dVdR}
\\[2mm]
\gamma_R(R)
&=&
\frac{\dot{P}_{\rm II}-\dot{P}_{\rm I}}{\dot{R}_{\rm I}-\dot{R}_{\rm II}}
+\frac{1}{2}\frac{d\mu}{dR}(\dot{R}_{\rm I}+\dot{R}_{\rm II}),
\label{Eq:gamma_R}
\end{eqnarray}
where the subscript I (II) indicates that those quantities are associated
with the TDHF trajectory I (II) at the collision energy $E_{\rm I}$
($E_{\rm II}=E_{\rm I}+\Delta E$). In the analysis given below
we set $\Delta E=0.01E_{\rm I}$. Note that all quantities on the right
hand side of Eqs.~(\ref{Eq:dVdR}) and (\ref{Eq:gamma_R}) should be
evaluated at the same relative distance, $R=R_{\rm I}=R_{\rm II}$.
The nucleus-nucleus potential $V_{\rm DD}(R)$ can be obtained by
a numerical integration of $dV_{\rm DD}/dR$. We refer to Refs.~\cite{
DDTDHF1,DDTDHF2,DDTDHF3} for details of numerical procedures.

\begin{figure} [t]
\includegraphics[width=8.6cm]{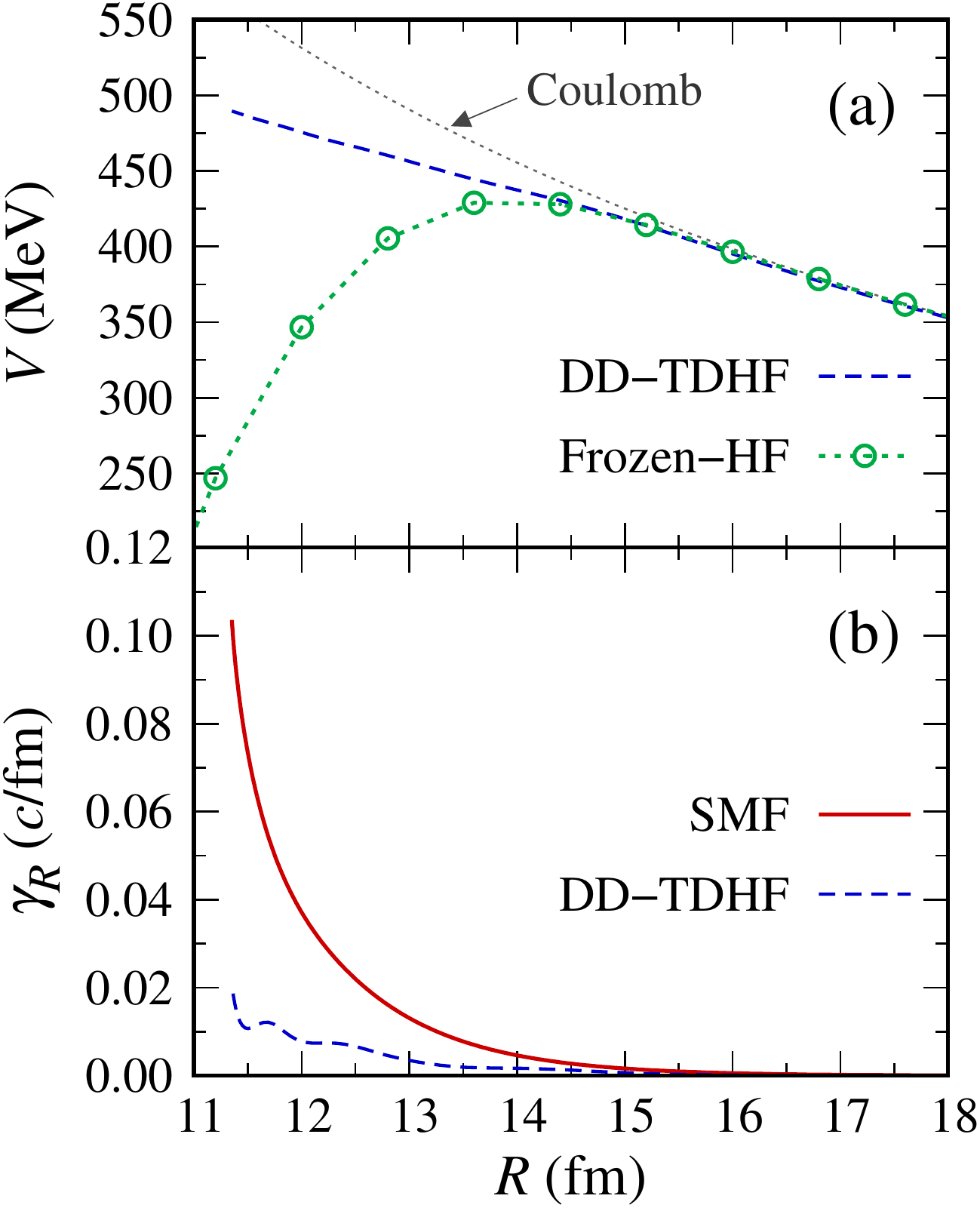}\vspace{-1mm}
\caption{
(a) The nucleus-nucleus potential $V(R)$ is shown as a function of
the relative distance $R$. The result obtained by the DD-TDHF method
is shown by a dashed line. For comparison, the nucleus-nucleus potential
obtained with the frozen-HF method and the point Coulomb potential are
also shown by open circles connected with dotted lines and a dotted line,
respectively. (b) The reduced radial friction coefficient $\gamma_R(R)$
for the incoming phase with $l=0$ is shown as a function of the relative
distance $R$. The result obtained with the SMF approach is shown by
a solid line, while that of DD-TDHF is shown by a dashed line.
}\vspace{-3mm}
\label{fig:C1}
\end{figure}

In Fig.~\ref{fig:C1}, the results obtained with the DD-TDHF method are presented.
In Fig.~\ref{fig:C1}(a), the extracted nucleus-nucleus potential $V_{\rm DD}(R)$
is shown as a function of the relative distance (dashed line), in comparison
with that obtained with the frozen-HF method \cite{Simenel(2017)}. In
the latter approach, the nucleus-nucleus potential is evaluated simply by
$V_{\rm FHF}(R)=E[\rho_{\rm P}+\rho_{\rm T}](R)-E[\rho_{\rm P}]-E[\rho_{\rm T}]$,
where $E[\rho]$ is the nuclear EDF and $\rho_{\rm P(T)}$ denotes the
ground-state density of the projectile (target) nucleus. It is to be reminded
that the Pauli exclusion principle among orbitals belonging to different nuclei
is neglected in the frozen-HF potential, which can be taken into account by
a density-constrained minimization technique \cite{Simenel(2017)}. Although
we should thus expect an increase of the potential at short distances
($R$\,$\lesssim$\,14.5\,fm), we present $V_{\rm FHF}$ to have an estimate
of the Coulomb barrier position. As can be seen from Fig.~\ref{fig:C1}(a), we observe
good agreement between $V_{\rm DD}$ and $V_{\rm FHF}$ at large distances
($R$\,$\gtrsim$\,15\,fm), as they should be. On the other hand, as the relative
distance decreases, we observe a monotonic increase of $V_{\rm DD}$, in contrast
to the significant reduction in $V_{\rm FHF}$. This is a characteristic behavior
observed for heavy systems, which is related to the fusion hindrance phenomenon
(see Ref.~\cite{DDTDHF3} for a detailed discussion).

In Fig.~\ref{fig:C1}(b), we show the reduced radial friction coefficient
$\gamma_R(R)$ for $l=0$ as a function of the relative distance. The result
based on the SMF approach is shown by a solid line, while the result of the
DD-TDHF method is shown by a dashed line. Since the mapping onto the
classical equations of motion breaks down around the turning point, we
compare the results for the incoming phase only. From the figure, we find that
the magnitude of the radial friction coefficient obtained with the SMF approach
is larger than the DD-TDHF result roughly by factor of 4. We note that the radial
friction coefficient in the SMF approach is consistent with the average TKEL in
TDHF according to Eq.~(\ref{AEq:E_diss}).

In the SMF approach the friction coefficient has a microscopic origin associated with
nucleon exchanges. On the other hand, in DD-TDHF the friction coefficient is
extracted based solely on the macroscopic relative motion of the colliding nuclei.
Based on the mapping of the classical equations of motion, the slowdown of
the relative motion after collision (due to incompressible character of nuclear
density and dinuclear shape formation) is mainly converted to the increase of the
nucleus-nucleus potential \cite{DDTDHF3}. However, the observed difference
between SMF and DD-TDHF approaches may indicate that energy dissipation
still takes place after the dinuclear system formation, because of the active
nucleon exchanges between the reactants. It would be interesting to reexamine
the origin of fusion hindrance in heavy systems based on the SMF approach.

\end{document}